\documentclass[showpacs,prl,aps,superscriptaddress,floatfix,tightenlines]{revtex4-1}

\usepackage{amsmath}
\usepackage{amssymb}
\usepackage{amsfonts}
\usepackage{epsfig}
\usepackage{subfigure}
\usepackage[dvips]{color}
\usepackage{bm}
\usepackage{amsthm}

\begin{document}

\title{Quantum-Atomic-Continuum-Coupled Model for Mechanical Behaviors in Micro-nano Simulations}

\author{Tiansi Han}

\email{hts@lsec.cc.ac.cn}

\author{Junzhi Cui}

\affiliation{LSEC, ICMSEC, Academy of Mathematics and Systems Science, CAS, Beijing, 100190, China}

\date{\today}

\begin{abstract}
For the numerical simulations of physical and mechanical behaviors of materials at the micro-nano scale, a coupled model with the effect of local quantum is presented in this paper. Unlike traditional methods, the transition region is not needed since the non-local mechanical effects and the constitutive relations are naturally involved by first principle density functional calculations. In order to identify and calculate the mechanical quantities at different scales, some necessary assumptions are made when solving Kohn-Sham equations. Basic deformation elements are introduced and mechanical tensors are explicitly derived based on the complex Bravais lattice. The responses of 3-demensional copper nanowires which composed of 25313 atoms under different external loads are simulated. Strain and stress fields are calculated and dislocation distributions are predicted during the damage process. Numerical results confirm the validity and transferability of this model.
\end{abstract}

\maketitle

\section{I. Introduction}

Computer simulations of physical and mechanical behaviors provide powerful tools for overcoming the difficulties that material properties cannot be designed. Generally, according to the scales, two well-established classes of models are widely used, i.e. molecular models and continuum models. The study of the relationships between structures and properties in various levels produces rigorous theoretical supports and effective computational techniques in new material developments (Ref. \cite{1}).

The molecular models, such as coarse grained molecular dynamics (CGMD)  (Ref. \cite{2}) and tight binding molecular dynamics (TBMD) (Ref. \cite{3}), provide detailed information about the micro-nano mechanism. But most of those rely heavily on the empirical potential functions between particles and grains. Thus it is still an open problem of formulating the multi-element systems (Ref. \cite{4}). Meanwhile, the density functional theory (DFT) is coded and applied to complex systems, and present papers confirm the physical and chemical quantities obtained by first principle calculations (Ref. \cite{5}). However, for the cases when numerous electrons contained and particular symmetry lacked, such as anisotropic films and polymer materials (Ref. \cite{6}), the local approximation fails in the descriptions of exchange and correlation effects. Note that even micro-nano devices are composed by more than ${10^3}$ atoms and electrons, thus high efficient computable models are urgent to be developed for micro-nano simulations.

The continuum models describe the physical responses of large scale systems and long running processes. But the sufficient accuracy initial parameters and boundary conditions are needed to make the results credible. So there emerged many multi-scale algorithms and models to solve the governing equations, such as the quasicontinuum method (QC) (Ref. \cite{7}), the bridging scale method (BSM) (Ref. \cite{8}), the heterogeneous multi-scale method (HMM) (Ref. \cite{9}), etc. Most of those models need special transition regions that link the atom and the continuum regions, while few of them link directly. Unfortunately, the boundary conditions under different scales are generally incompatible, thus occurs the ``ghost force'' (Ref. \cite{10}). In addition, the transition regions are not easy to determine in real-time simulations.

For approaching the real physical processes in the macro-meso-micro coupled simulations, a Quantum-Atomic-Continuum-Coupled method (QACC) with the local quantum effects considered is modeled. In QACC model, the atom and the continuum regions are linked without the transition regions. A linear expansion method is employed and the high-order terms are truncated when solving the electronic structure Kohn-Sham equation. Unlike traditional finite element triangulations (Ref. \cite{11}), we use primitive cells of the complex Bravais lattice to construct basic deformation elements and deduce the mechanical tensors, as most of the polycrystalline systems and the micro-nano devices have particular topological structures. Furthermore, the physical properties of micro-nano materials are extremely sensitive to the variation of intrinsic structures (Ref. \cite{12}), and the material damages occur exactly in the defect areas where the micro-structures deconstruct. Thus our model may give better results of plastic variations and constitutive parameters.

The remainder of this paper is outlined as follows: In Section II, a quantum energy density distribution of large atomic system is described. Section III introduces basic deformation elements of the complex Bravais lattice firstly, and then the explicit expressions of strain, stress and other mechanical tensors are derived. Section IV shows the numerical simulations of copper nanowires composed of 25313 atoms under tension and bending. Strain and stress fields are calculated to confirm our model. Conclusions are summarized in Section V.

\section{II. Quantum Energy Density of Atomic System}

As mentioned previously, it takes enormous amount of efforts solving the atomic system completely by first principle calculations. For the computations of multi-scale and multi-model coupled quantities, some necessary assumptions and techniques are adopted in this section. Keeping certain precision, QACC model can simulate damage processes and behaviors of materials at the micro-nano scale by the ab initio quantum mechanics.

Consider a local atomic system which composed with $N_{nucl}$ nuclei and $n_{elec}$ electrons. Here the uppercase letters are used for quantities of nuclei and the lowercase letters for electrons. The Hartree atomic units are adopted so that the four fundamental physical constants are unity, i.e. ${m_e} = e = 4\pi {\varepsilon _0} = \hbar = 1$. According to density functional theory, the non-relativistic Kohn-Sham equation can be written as
\begin{eqnarray}
\left( { - \frac{1}{2}\nabla _i^2 + V\left( \rho \right) + \int {dr\frac{{\rho \left( {r'} \right)}}{{| {r - r'} |}}} + \frac{{\delta {E_{XC}}\left( \rho \right)}}{{\delta \rho }}} \right){\phi _i} = {\varepsilon _i}{\phi _i} .
\label{eq:1}
\end{eqnarray}
Note that ${{E}_{xc}}\left( \rho \right)$ is the exchange-correlation energy, $\rho \left( r \right)$ is the electron density, $V\left( \rho \right)$ contains both the external field potential ${V_{ext}}\left( \rho \right)$ and the Coulomb interaction $- \sum\limits_I {\frac{{{Z_I}}}{{| {{r_i} - {R_I}} |}}}$, and ${Z_I}$ is the atomic number (Ref. \cite{13}). Thus the ground state total energy is
\begin{eqnarray}
{E_{total}} && = \sum\limits_i {{\varepsilon _i}} - \iint {drdr'\frac{1}{2}\frac{{\rho \left( r \right)\rho \left( {r'} \right)}}{{| {r - r'} |}}} - \int {dr\frac{{\delta {E_{XC}}\left( \rho \right)}}{{\delta \rho }}\rho } + {E_{XC}}\left( \rho \right) + {E_{nucl}} \nonumber \\
&& = \sum\limits_i {{\varepsilon _i}} + {E_{rep}} + {E_{nucl}} .
\label{eq:2}
\end{eqnarray}

Now we consider ${E_{total}}$ as contributions of each atom centered at ${R_I}$, a traditional consideration is the average distribution, and each atom occupies ${E^I} = \frac{{{E_{total}}}}{{{N_{nucl}}}}$ of the total energy. But when a micro-damage occurs in materials, stress concentrates, crack propagates, and finally energy dissipates in the defect area. Thus the simple average method is not able to reflect the energy variation between atoms in and around the defect area (Ref. \cite{14}). So it is reasonable to be deliberate when large deformations or chemical reactions occur to the micro-structure. Suppose that the initial electron density is the superposition of electron densities around each atom
\begin{eqnarray}
\rho \left( r \right) = \sum\limits_I {{\rho _I}\left( r \right)} ,
\label{eq:3}
\end{eqnarray}
where
\begin{eqnarray}
{\rho _I}\left( r \right) = \sum\limits_{i \in I} {{{| {{\phi _i}\left( r \right)} |}^2}} ,
\label{eq:4}
\end{eqnarray}
and $i \in I$ means to sum over all electrons around atom $I$. We formulate Eq. \eqref{eq:2} as
\begin{eqnarray}
\sum\limits_{i}{{{\varepsilon }_{i}}} = && \sum\limits_{I}{\varepsilon {}_{I}}+\frac{1}{2}\sum\limits_{I}{\sum\limits_{J\ne I}{\left[ {{\varepsilon }_{I+J}} - {{\varepsilon }_{I}} - {{\varepsilon }_{J}} \right]}} + \frac{1}{6}\sum\limits_{I}{\sum\limits_{J\ne I}{\sum\limits_{\begin{smallmatrix} K\ne I \\ K\ne J
\end{smallmatrix}}{\left[ {{\varepsilon }_{I + J + K}} \right.}}} \nonumber \\
&& \left. { - a {\varepsilon _{I + J}} - b {\varepsilon _{J + K}} - c {\varepsilon _{I + K}} + {\varepsilon _I} + {\varepsilon _J} + {\varepsilon _k}} \right] + \cdots ,
\label{eq:5}
\end{eqnarray}
\begin{eqnarray}
{E_{rep}} = && \sum\limits_I {{E_{rep}}\left( {{\rho _I}} \right)} + \frac{1}{2}\sum\limits_I {\sum\limits_{J \ne I} {\left[ {{E_{rep}}\left( {{\rho _I} + {\rho _J}} \right) - {E_{rep}}\left( {{\rho _I}} \right) - {E_{rep}}\left( {{\rho _J}} \right)} \right]} } \nonumber \\
&& + \frac{1}{6}\sum\limits_{I}{\sum\limits_{J\ne I}{\sum\limits_{\begin{smallmatrix} K\ne I \\ K\ne J
\end{smallmatrix}}{\left[ {{E}_{rep}}\left( {{\rho }_{I}} + {{\rho }_{J}} + {{\rho }_{K}} \right) - a {{E}_{rep}}\left( {{\rho }_{I}} + {{\rho }_{J}} \right) - b {{E}_{rep}}\left( {{\rho }_{J}} + {{\rho }_{K}} \right) \right.}}} \nonumber \\
&& \left. { - c {E_{rep}}\left( {{\rho _I} + {\rho _K}} \right) + {E_{rep}}\left( {{\rho _I}} \right) + {E_{rep}}\left( {{\rho _J}} \right) + {E_{rep}}\left( {{\rho _K}} \right)} \right] + \cdots ,
\label{eq:6}
\end{eqnarray}
where
\begin{eqnarray}
{\varepsilon _I} = \sum\limits_{i \in I} {{\varepsilon _i}} = \sum\limits_{i \in I} {\langle {{\phi _i}} | - \frac{1}{2}\nabla _i^2 + V\left( {{\rho _I}} \right) + \int {dr\frac{{{\rho _I}\left( {r'} \right)}}{{| {r - r'} |}}} + \frac{{\delta {E_{XC}}\left( {{\rho _I}} \right)}}{{\delta {\rho _I}}} | {{\phi _i}} \rangle } ,
\label{eq:7}
\end{eqnarray}
and
\begin{eqnarray}
{E_{rep}}\left( {{\rho _I}} \right) = - \iint {drdr'\frac{1}{2}\frac{{{\rho _I}\left( r \right){\rho _I}\left( {r'} \right)}}{{| {r - r'} |}}} - \int {dr\frac{{\delta {E_{XC}}\left( {{\rho _I}} \right)}}{{\delta \rho }}{\rho _I}} + {E_{XC}}\left( {{\rho _I}} \right) .
\label{eq:8}
\end{eqnarray}
$a, b, c$ in Eq. \eqref{eq:5} and Eq. \eqref{eq:6} can be acquired by the undetermined coefficient method.

Actually, it is a linear expansion of the complicated ${E_{xc}}\left( \rho \right)$ and the highly non-localized integral $\int {dr\frac{{\rho \left( {r'} \right)}}{{| {r - r'} |}}}$, and an approach to the initial many-body problem with linear combinations of small-case many-body problems. In physics, it is an approximation of the long-range interaction with short-range interactions. If only two-center contributions considered, We can finally deduce an atomic-center energy distribution pattern, and rewrite Eq. \eqref{eq:2} with Greek letters
\begin{eqnarray}
{E_{total}} = \sum\limits_\alpha {{E^\alpha }} ,
\label{eq:9}
\end{eqnarray}
where
\begin{eqnarray}
{E^\alpha } = {\varepsilon ^\alpha } + E_{rep}^\alpha + E_{nucl}^\alpha ,
\label{eq:10}
\end{eqnarray}
\begin{eqnarray}
{\varepsilon ^\alpha } = {\varepsilon _\alpha } + \frac{1}{2}\sum\limits_{\beta \ne \alpha } {\left[ {{\varepsilon _{\alpha + \beta }} - {\varepsilon _\alpha } - {\varepsilon _\beta }} \right]} ,
\label{eq:11}
\end{eqnarray}
\begin{eqnarray}
E_{rep}^\alpha = {E_{rep}}\left( {{\rho _\alpha }} \right) + \frac{1}{2}\sum\limits_{J \ne I} {\left[ {{E_{rep}}\left( {{\rho _{\alpha + \beta }}} \right) - {E_{rep}}\left( {{\rho _\alpha }} \right) - {E_{rep}}\left( {{\rho _\beta }} \right)} \right]} ,
\label{eq:12}
\end{eqnarray}
\begin{eqnarray}
E_{nucl}^\alpha = \frac{1}{2}{M_\alpha }\dot R_\alpha ^2 + \sum\limits_{\beta \ne \alpha } {\frac{1}{2}\frac{{{Z_\alpha }{Z_\beta }}}{{| {{R_\alpha } - {R_\beta }} |}}} .
\label{eq:13}
\end{eqnarray}

\begin{figure}[ht]
\subfigure[]{\label{fig:1:a}\includegraphics[width=2.5in]{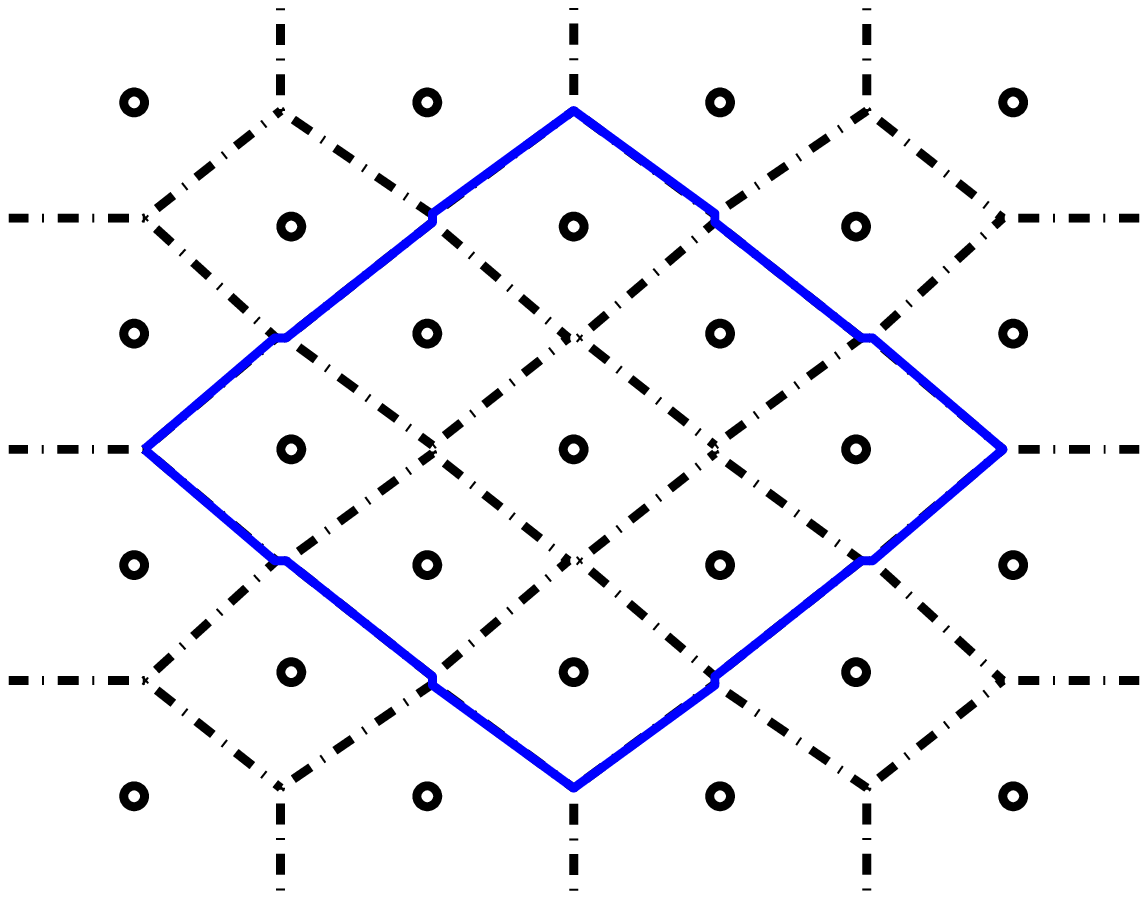}}
\subfigure[]{\label{fig:1:b}\includegraphics[width=2.5in]{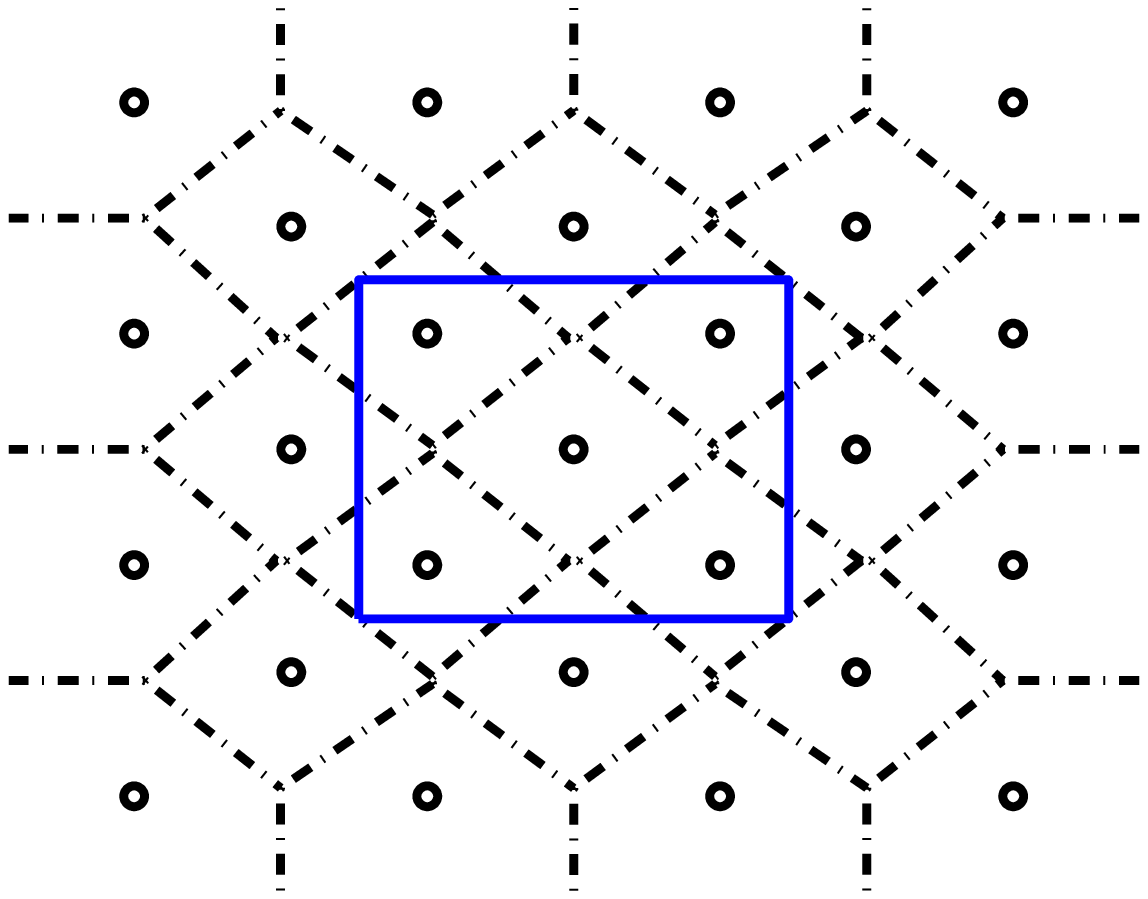}}
\caption{The current configuration contains total (a) and part (b) of Wigner-Seitz cells. Solid lines are used for the current configuration, and dash lines are for the Wigner-Seitz cells (color on line).}
\label{fig:1}
\end{figure}

We use the volumes of Wigner-Seitz cells as the energy density is a volume average quantity. Note that the possession ratio is needed if the current configuration is not a standard Wigner-Seitz cell, as seen in FIG. \ref{fig:1}.
The quantum strain energy density of atomic system, i.e. strain energy per unit reference configuration volume, is then given by
\begin{eqnarray}
{w_r} = \frac{1}{V}\sum\limits_{\alpha \in {\Omega _r}} {{\eta _\alpha }{E^\alpha }} ,
\label{eq:14}
\end{eqnarray}
and $\alpha \in {\Omega _r}$ means to sum over all atoms in the current configuration ${\Omega _r}$. There are several ways to determine possession ratios ${\eta _\alpha }$: If ${E^\alpha }$ is uniform in the cell region ${C_\alpha }$, ${\eta _\alpha } = \frac{{| {{C_\alpha } \cap V} |}}{{| {{C_\alpha }} |}}$ is adopted; Otherwise, ${\eta _\alpha }$ is also relevant to the gradient of electron density in ${\Omega _r}$, and we formulate ${\eta '_\alpha } = {\eta '_\alpha }\left( {{\eta _\alpha },{\rho _\alpha },\nabla {\rho _\alpha }} \right)$ as similar to the generalized gradient approximation (GGA) in DFT (Ref. \cite{15}). In this paper, we take the former possession ratio for simplicity.

The free energy density ${a_r}$ of atomic system in the current configuration is obtained with an additional entropy term to Eq. \eqref{eq:14}. Following Tadmor's work (Ref. \cite{16}), it is denoted by
\begin{eqnarray}
{a_r} = {w_r} - T{s_r} = \frac{1}{V}\sum\limits_{\alpha \in {\Omega _r}} {\left\{ {{\eta _\alpha }{E^\alpha } + {k_B}T\sum\limits_{i \in \alpha } {\left[ {{f_i}\ln {f_i} + \left( {1 - {f_i}} \right)\ln \left( {1 - {f_i}} \right)} \right]} } \right\}} ,
\label{eq:15}
\end{eqnarray}
where ${s_r}$ is the entropy density, ${f_i} = \frac{1}{{\exp \left( {{{{\varepsilon _i} - {\varepsilon _f}} \mathord{\left/ {\vphantom {{{\varepsilon _i} - {\varepsilon _f}} {{k_B}T}}} \right. \kern-\nulldelimiterspace} {{k_B}T}}} \right) + 1}}$ is the Fermi function, ${\varepsilon _f}$ is the Fermi energy and ${k_B}$ is the Boltzmann constant. Till now we have determined the basic features of QACC model at the quantum-atomic scale.

\section{III. Deformation Environment Framework}

\begin{figure}[ht]
\subfigure[]{\label{fig:2:a}\includegraphics[width=2.0in]{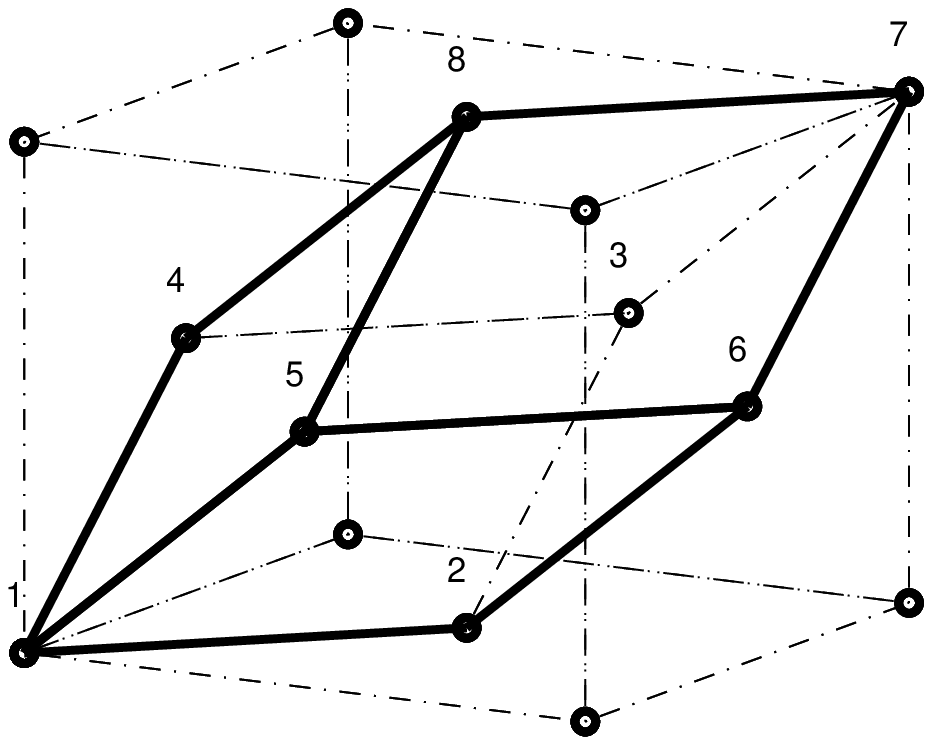}}
\subfigure[]{\label{fig:2:b}\includegraphics[width=2.0in]{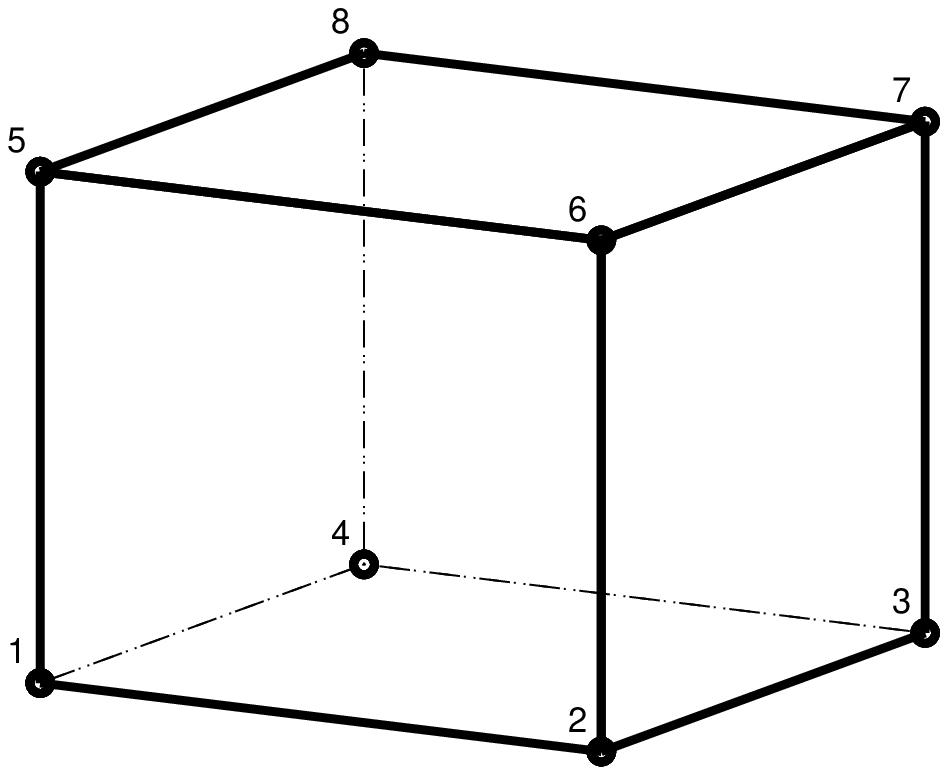}}
\subfigure[]{\label{fig:2:c}\includegraphics[width=2.0in]{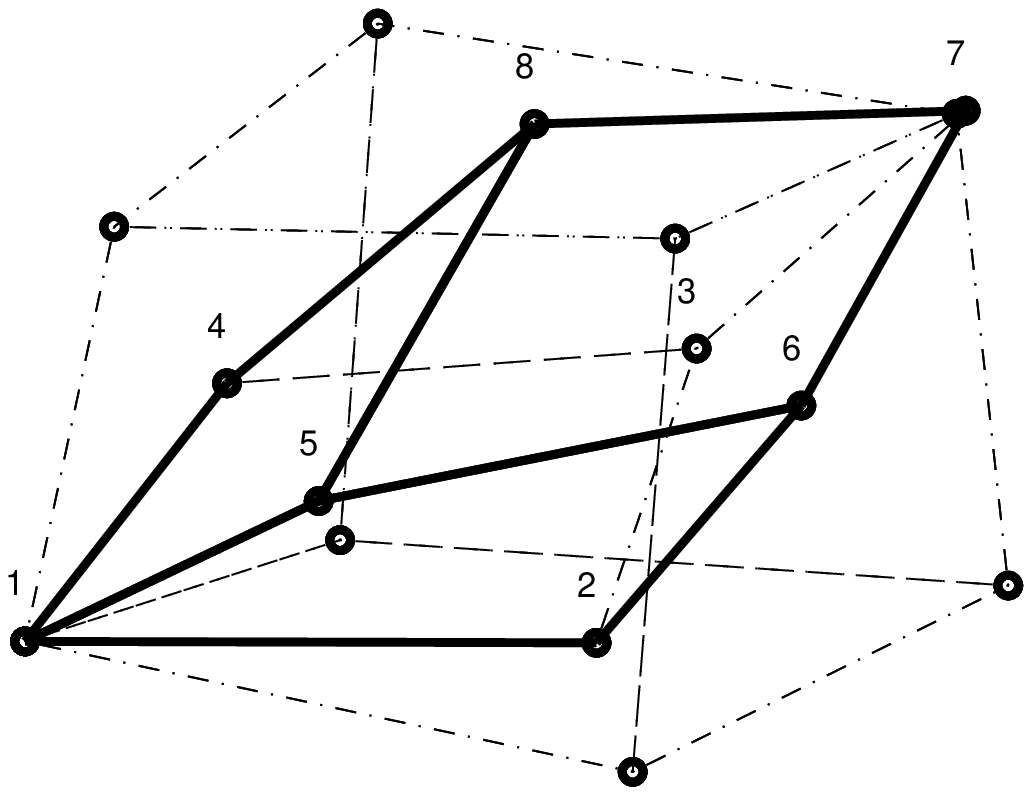}}
\caption{The primitive cell of the FCC structure in the reference configuration with the vertexes 1 to 8 (a) is firstly transformed to a standard cubic (b), and then mapped to the current configuration (c).}
\label{fig:2}
\end{figure}

In order to simulate the process from defects to damages of micro-nano devices, and to formulate the constitutive law of materials at the micro-nano scale, the total displacement field of each atom is decomposed into a low oscillatory deformation part and a high frequent vibration part ${X_{total}} = X + {X_{vib}}$, as the boundary conditions at the atom scale and the continuum scale differ by orders of magnitudes. Here the deformation part is obtained by the average position of 2 picoseconds of each atom in each loading step. And we use the structure of primitive cells of the complex Bravais lattice as basic deformation elements in QACC model. In the continuum framework, consider a one-to-one mapping $\varphi$ from the reference configuration ${\Omega _0}$ to the current configuration ${\Omega _r}$, and for every $X \in {\Omega _0}$, $x = \varphi \left( X \right) \in {\Omega _r}$. Suppose that the deformation environment of point $X$ at the continuum scale is uniform with that of a volume element centered at $X$ at the atomic scale. Also suppose that the electrons of the lattice are adiabatic kept up with the nuclei under deformations. Using the face-centered-cubic (FCC) structure as an example, we can derive the deformation gradient as follows:

(i) as seen in FIG. \ref{fig:2}, transform the primitive cell in the reference configuration with its coordinates $X \in {\Omega _0}$ to a standard hexahedron configuration $\zeta \in {\left[ { - 1,1} \right]^3}$ with a vertex-to-vertex transformation tensor $T$. Thus
\begin{eqnarray}
\zeta = T\left( {X - {X^{Center}}} \right) ,
\label{eq:16}
\end{eqnarray}
where $T = \frac{2}{{lc}}\left( {\begin{array}{*{20}{c}}1&1&{ - 1} \\{ - 1}&1&{ - 1} \\1&1&1\end{array}} \right)$ is the transformation tensor, $lc$ is the lattice constant, and ${X^{Center}}$ is the geometric center of the primitive cell.

(ii) The current configuration is traditionally obtained by MD simulations. With the help of iso-parametric finite element method skills (Ref. \cite{17}), the standard hexahedron configuration is transformed to the current configuration with another vertex-to-vertex transformation tensor ${N_p}$. Here the superscripts are used for the coordinate components and the subscripts are for vertexes. For any $x \in {\Omega _r}$, the coordinate components of $x$, are obtained by
\begin{eqnarray}
{x^i} = \sum\limits_{p = 1}^8 {{N_p}\left( {{\zeta ^1},{\zeta ^2},{\zeta ^3}} \right)x_p^i} ,
\label{eq:17}
\end{eqnarray}
where $x_p^i$ are the coordinate components of the vertexes 1 to 8, and the interpolation functions are
\begin{eqnarray}
{{N}_{p}}\left( {{\zeta }^{1}},{{\zeta }^{2}},{{\zeta }^{3}} \right) = \frac{1}{8}\prod\limits_{\begin{smallmatrix} q = 1 \\ q\ne k
\end{smallmatrix}}^{3}{\left( 1 + \zeta _{p}^{q}{{\zeta }^{k}} \right)} ,
\label{eq:18}
\end{eqnarray}
and $\left( {\zeta _p^1,\zeta _p^2,\zeta _p^3} \right) = \left( { \pm 1, \pm 1, \pm 1} \right)$.

(iii) Finally, by Eq. \eqref{eq:16}, Eq. \eqref{eq:17} and Eq. \eqref{eq:18}, the deformation gradient tensor is derived explicitly
\begin{eqnarray}
{F_{iJ}} && = \frac{{\partial \varphi \left( {{X_I}} \right)}}{{\partial {X_J}}} = \frac{{\partial {x_i}}}{{\partial {X_J}}} = \sum\limits_{k = 1}^3 {\sum\limits_{p = 1}^8 {\frac{{\partial {x_i}}}{{\partial {N_p}}}\frac{{\partial {N_p}}}{{\partial {\zeta ^k}}}\frac{{\partial {\zeta ^k}}}{{\partial {X_J}}}} } \nonumber \\
&& = \frac{1}{8}\sum\limits_{k = 1}^{3}{\left\{ \sum\limits_{p = 1}^{8}{\left[ \prod\limits_{\begin{smallmatrix} q = 1 \\ q\ne k \end{smallmatrix}}^{3}{\left( 1+\zeta _{p}^{q}{{T}^{q}}\left( X-{{X}^{C}} \right) \right)} \right]}\zeta _{p}^{k}x_{p}^{i} \right\}{{T}^{kJ}}} .
\label{eq:19}
\end{eqnarray}
And in the current configuration the distance between any two points ${x_\alpha }$ and ${x_\beta }$ of the local atomic system at the continuum scale is
\begin{eqnarray}
r_{\alpha \beta }^i && = x_\beta ^i - x_\alpha ^i = \varphi \left( {X_\beta ^I} \right) - \varphi \left( {X_\alpha ^I} \right) = \int_0^1 {\frac{{\partial \varphi \left( {X_\alpha ^I - s\left( {X_\beta ^I - X_\alpha ^I} \right)} \right)}}{{\partial s}}ds} \nonumber \\
&& = \int_0^1 {F\left( {{X_\alpha } - s\left( {{X_\beta } - {X_\alpha }} \right)} \right)\left( {{X_\beta } - {X_\alpha }} \right)ds} = \int_0^1 {F\left( {{X_\alpha } - s{R_{\alpha \beta }}} \right){R_{\alpha \beta }}ds} .
\label{eq:20}
\end{eqnarray}
Remark that   is just the deformation part of the total displacement field. Furthermore, the right Cauchy strain tensor is ${{\rm E}_{IJ}} = \sum\limits_{k = 1}^3 {{F_{kI}}{F_{kJ}}}$. And for more examples of other primitive cell structures, see J. Cui' work (Ref. \cite{18}).

The first Piola-Kirchhoff stress tensor and the elastic constant tensor of atomic system, which denoted by the first and second partial derivatives of the free energy function with respect to the deformation gradient, are derived by
\begin{eqnarray}
{P_{hK}} = && \frac{1}{V}\sum\limits_{\alpha \in {\Omega _r}} {\left\{ {\sum\limits_{\beta \ne \alpha } {{\eta _\alpha }\left[ {\int_0^1 {\left( {\frac{{\partial \left( {{\varepsilon ^\alpha }\left( r \right) + E_{rep}^\alpha \left( \rho \right) + {Z_\alpha }{Z_\beta }{r^{ - 1}}} \right)}}{{2\partial r}}} \right)_{r = \int_0^1 {F\left( {X + \left( {s' - s} \right){R_{\alpha \beta }}} \right){R_{\alpha \beta }}ds'} }^hR_{\alpha \beta }^Kds} } \right]} } \right.} \nonumber \\
&& \left. { - \sum\limits_{i \in \alpha } {\left[ {\frac{{{g_i}}}{{{k_B}T}}\left( {{\varepsilon _i} - \frac{{\sum\limits_m {{g_m}{\varepsilon _m}} }}{{\sum\limits_m {{g_m}} }}} \right)\int_0^1 {\left( {\frac{{\partial {\varepsilon _i}\left( r \right)}}{{\partial r}}} \right)_{r = \int_0^1 {F\left( {X + \left( {s' - s} \right){R_{\alpha \beta }}} \right){R_{\alpha \beta }}ds'} }^hR_{\alpha \beta }^Kds} } \right]} } \right\} ,
\label{eq:21}
\end{eqnarray}
and
\begin{eqnarray}
{C_{IJKL}} = 2\sum\limits_{h,m,N} {\left[ {\frac{{\partial {{\left( {{F^{ - 1}}} \right)}_{Ih}}}}{{\partial {F_{mN}}}}{P_{hJ}} + {{\left( {{F^{ - 1}}} \right)}_{Ih}}{D_{hJmN}}} \right]\frac{{\partial {F_{mN}}}}{{\partial {E_{KL}}}}} .
\label{eq:22}
\end{eqnarray}
Here $\frac{{\partial {\varepsilon _i}}}{{\partial r}} = \langle {{\phi _i}} | \frac{{\partial \left( { - \frac{1}{2}\nabla _i^2 + {V_{eff}}\left( {{\rho _\alpha }} \right)} \right)}}{{\partial r}} | {{\phi _i}} \rangle$, are obtained by Hellmann-Feynman theorem (Ref. \cite{19}), ${D_{iJkL}} = \frac{{{\partial ^2}{a_r}}}{{\partial {F_{iJ}}\partial {F_{kL}}}} = \frac{{\partial {P_{iJ}}}}{{\partial {F_{kL}}}}$ and $\frac{{\partial F}}{{\partial E}} = {\left( {\frac{{\partial E}}{{\partial F}}} \right)^{ - 1}}$. The detailed derivation of Eq. \eqref{eq:21} and Eq. \eqref{eq:22} are given in the Appendix. In addition, the Cauchy stress tensor is
\begin{eqnarray}
\sigma = \frac{1}{{\det \left( F \right)}}P{F^T} = \frac{1}{{2\det \left( F \right)}}\left( {P{F^T} + F{P^T}} \right) .
\label{eq:23}
\end{eqnarray}

From Eq. \eqref{eq:21}, it is easy to obtain a corollary as follows: when the current configuration is taken as the reference configuration, the deformation gradient degenerates into the identity. Thus both the first Piola-Kirchhoff stress and the Cauchy stress can be written as the Virial stress formulation
\begin{eqnarray}
{\tau _{hK}} = && \frac{1}{V}\sum\limits_{\alpha \in {\Omega _r}} {\left\{ {\sum\limits_{\beta \ne \alpha } {{\eta _\alpha }\left[ {{{\left( {\frac{{\partial \left( {{\varepsilon ^\alpha }\left( {{r_{\alpha \beta }}} \right) + E_{rep}^\alpha \left( \rho \right) + {Z_\alpha }{Z_\beta }r_{\alpha \beta }^{ - 1}} \right)}}{{2\partial r}}} \right)}^h}r_{\alpha \beta }^K} \right]} } \right.} \nonumber \\
&& \left. { - \sum\limits_{i \in \alpha } {\left[ {\frac{{{g_i}}}{{{k_B}T}}\left( {{\varepsilon _i} - \frac{{\sum\limits_m {{g_m}{\varepsilon _m}} }}{{\sum\limits_m {{g_m}} }}} \right){{\left( {\frac{{\partial {\varepsilon _i}\left( {{r_{\alpha \beta }}} \right)}}{{\partial r}}} \right)}^h}r_{\alpha \beta }^K} \right]} } \right\} .
\label{eq:24}
\end{eqnarray}
The Virial stress is not able to reflect thermo-mechanical effects of atomic system because of neglecting the deformation behaviors (Ref. \cite{20}).

The framework of QACC model has now been built completely. The mechanical quantities of large atomic system can be calculated by ab initio quantum mechanics and also the physical properties of micro-nano materials can be simulated. The model partly overcomes the difficulties in solving large scale systems completely by first principle calculations. With no need of empirical potential functions, QACC model guarantees better accuracy and broader use in numerical simulations.

\section{IV. Numerical Simulations}

The single-crystal copper nanowires composed of 25313 atoms are constructed with the size $43.38 \times 144.6 \times 43.38{A^3}$. And the responses of the $\left\{ {100} \right\}\left\langle {001} \right\rangle$ nanowires under external tension and bending at $300K$ are studied. Dislocation distributions are shown while strain and stress fields are calculated. Simulations have been carried out by using Gaussian09 programs (Ref. \cite{21}). The pseudo-potential basis set of LanL2DZ (Ref. \cite{22}) is employed and the exchange-correlation energy functional is modeled with TPSS (Ref. \cite{23}). Relaxation procedures with 60 picoseconds are applied after the initial configuration and every external loading step are handled.

\subsection{TENSION}

\begin{figure}[ht]
\subfigure[]{\label{fig:3:a}\includegraphics[width=2.5in]{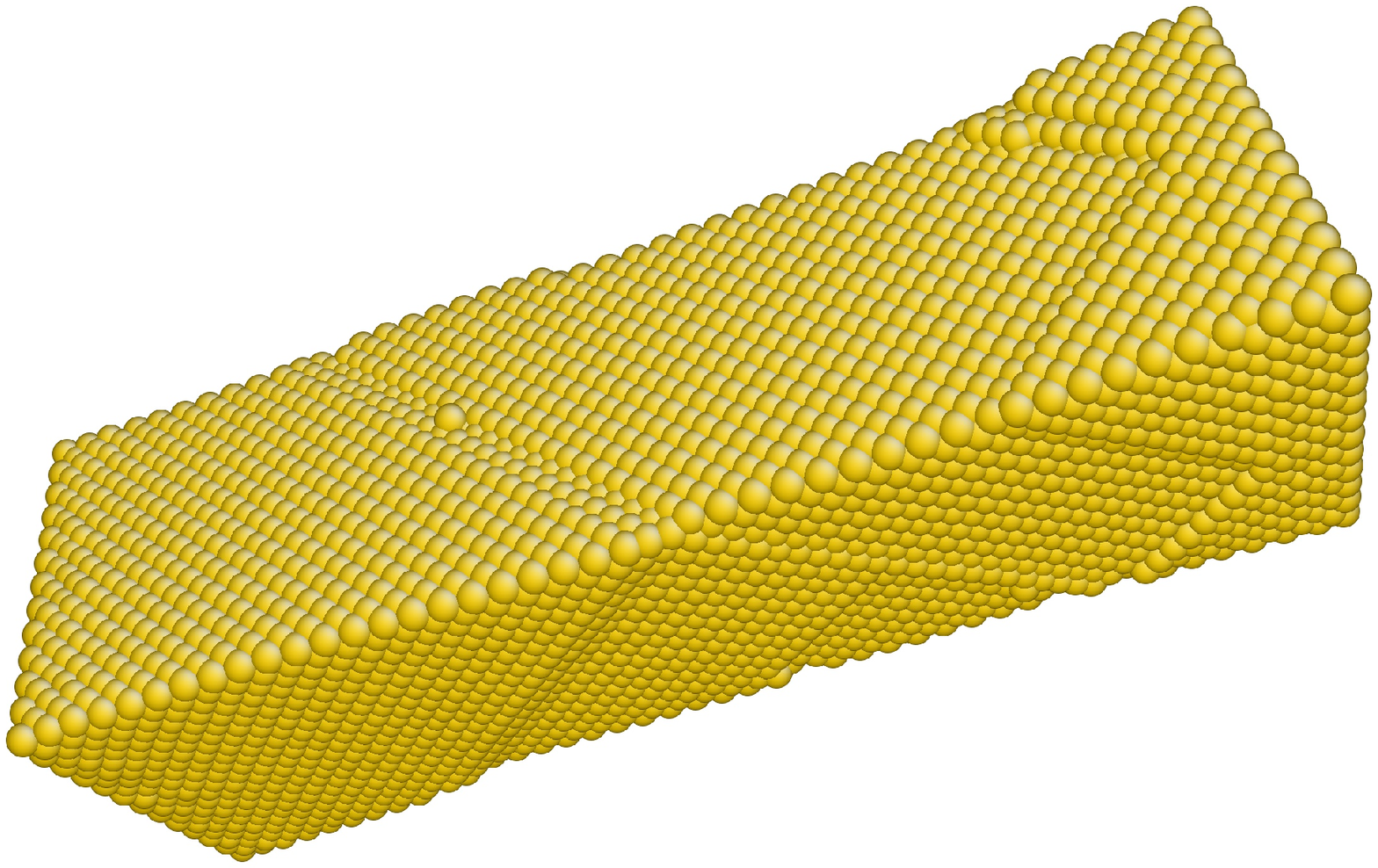}}
\subfigure[]{\label{fig:3:b}\includegraphics[width=2.5in]{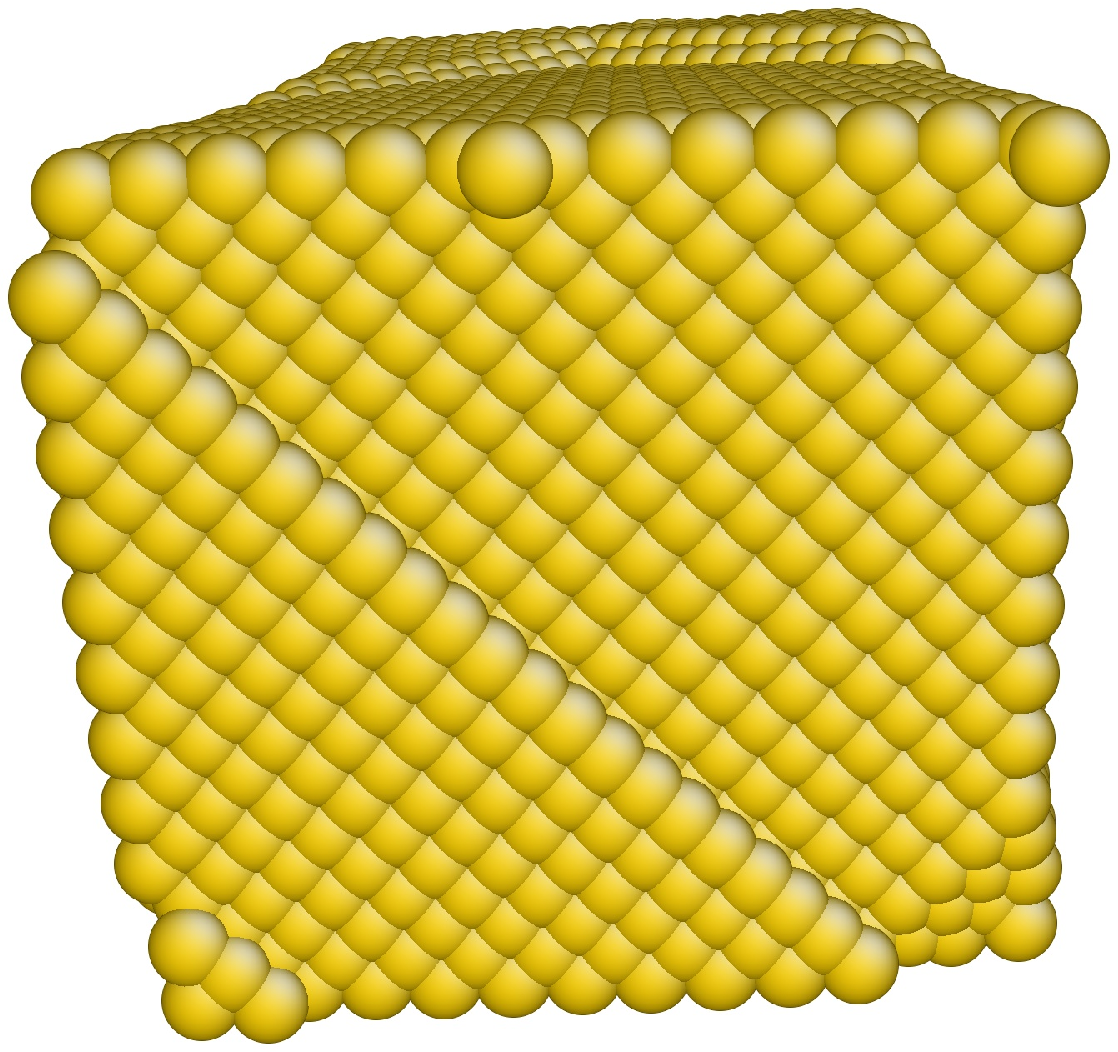}}
\caption{The configuration and the dislocation distribution of the nanowire (a) and the cross section (b) under tension at $3600$ picoseconds (color on line).}
\label{fig:3}
\end{figure}

\begin{figure}[ht]
\subfigure[]{\label{fig:4:a}\includegraphics[width=2.5in]{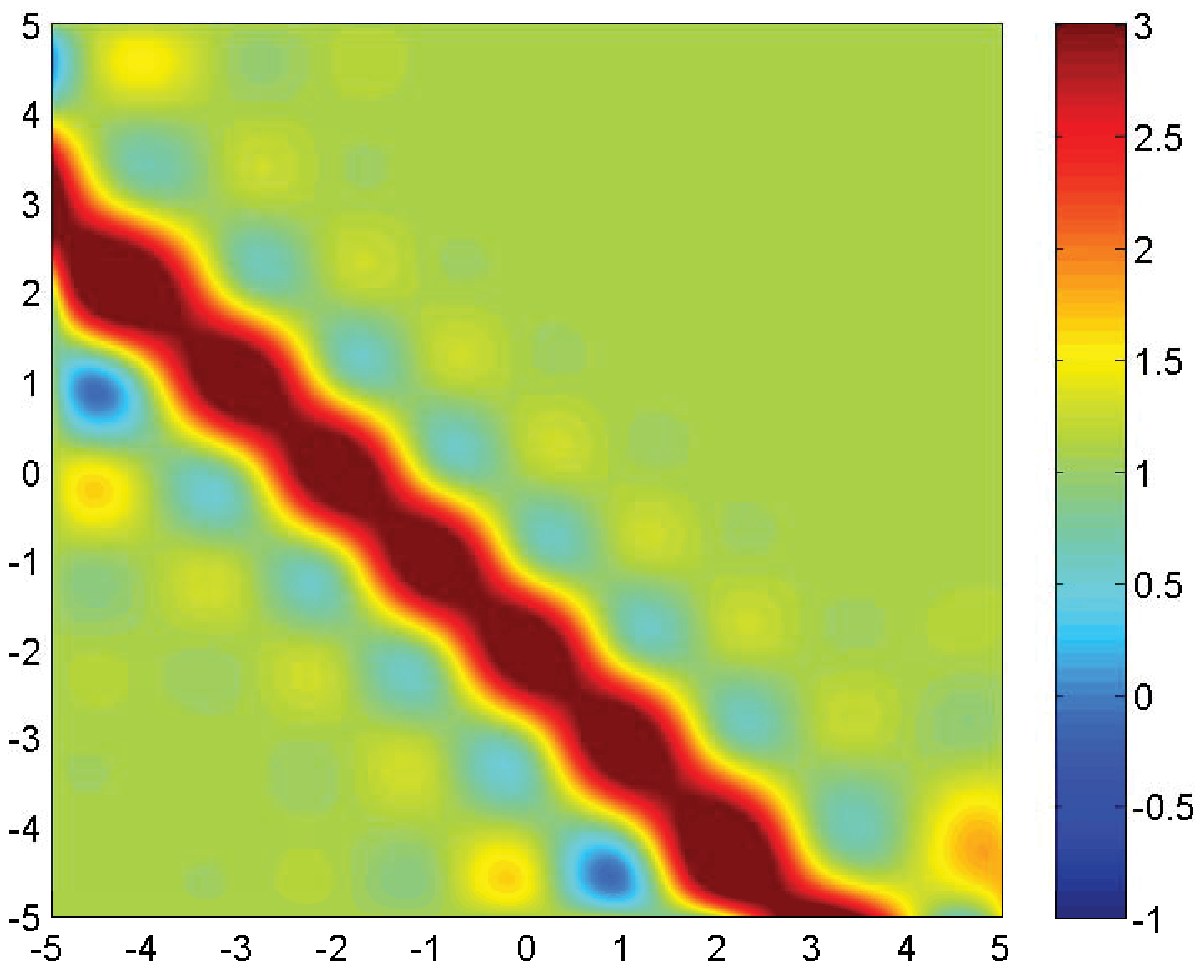}}
\subfigure[]{\label{fig:4:b}\includegraphics[width=2.5in]{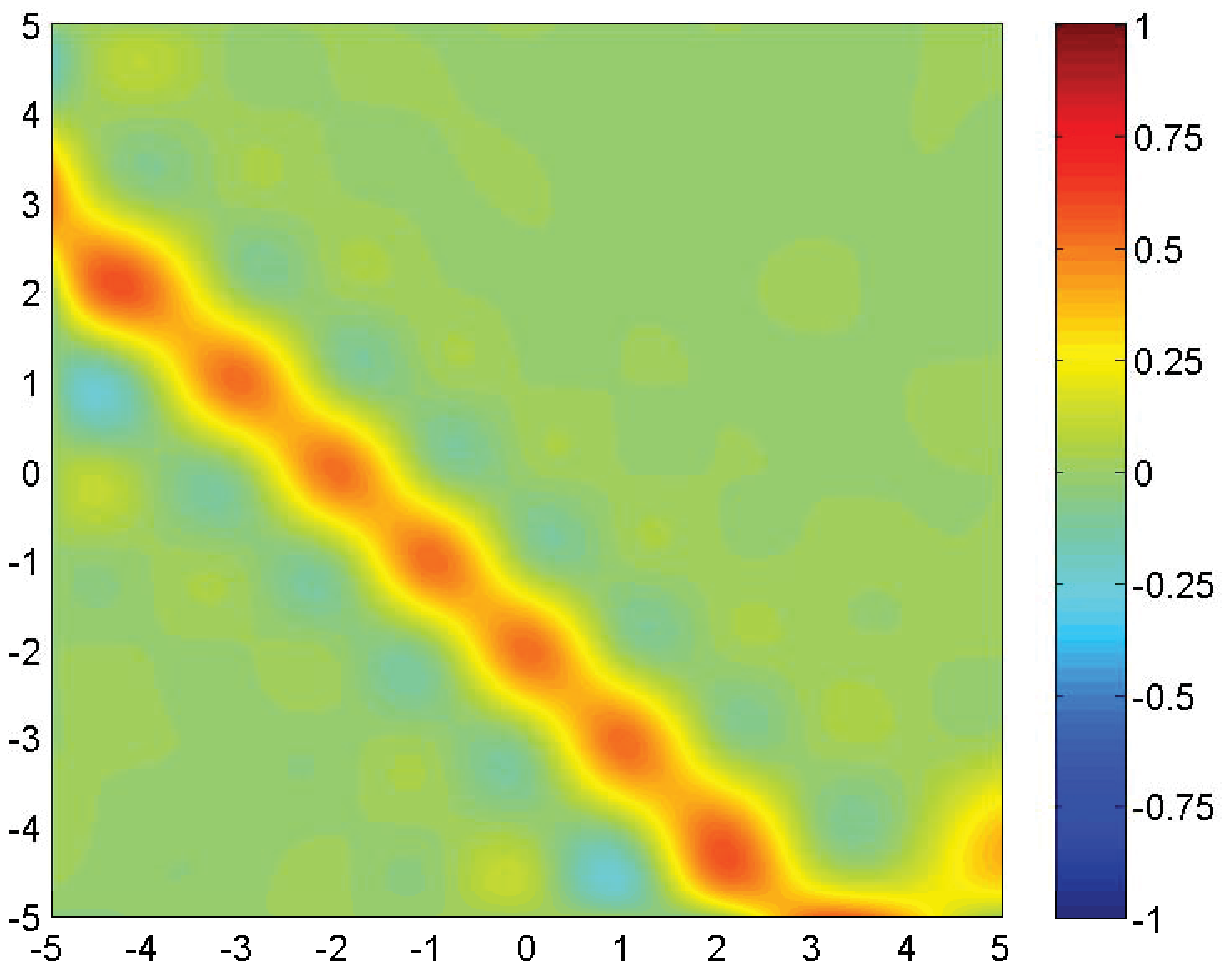}}
\caption{The Cauchy strain component ${E_{yy}}$ (a), ${E_{xz}}$ (b) under tension (color on line).}
\label{fig:4}
\end{figure}

\begin{figure}[ht]
\subfigure[]{\label{fig:5:a}\includegraphics[width=2.5in]{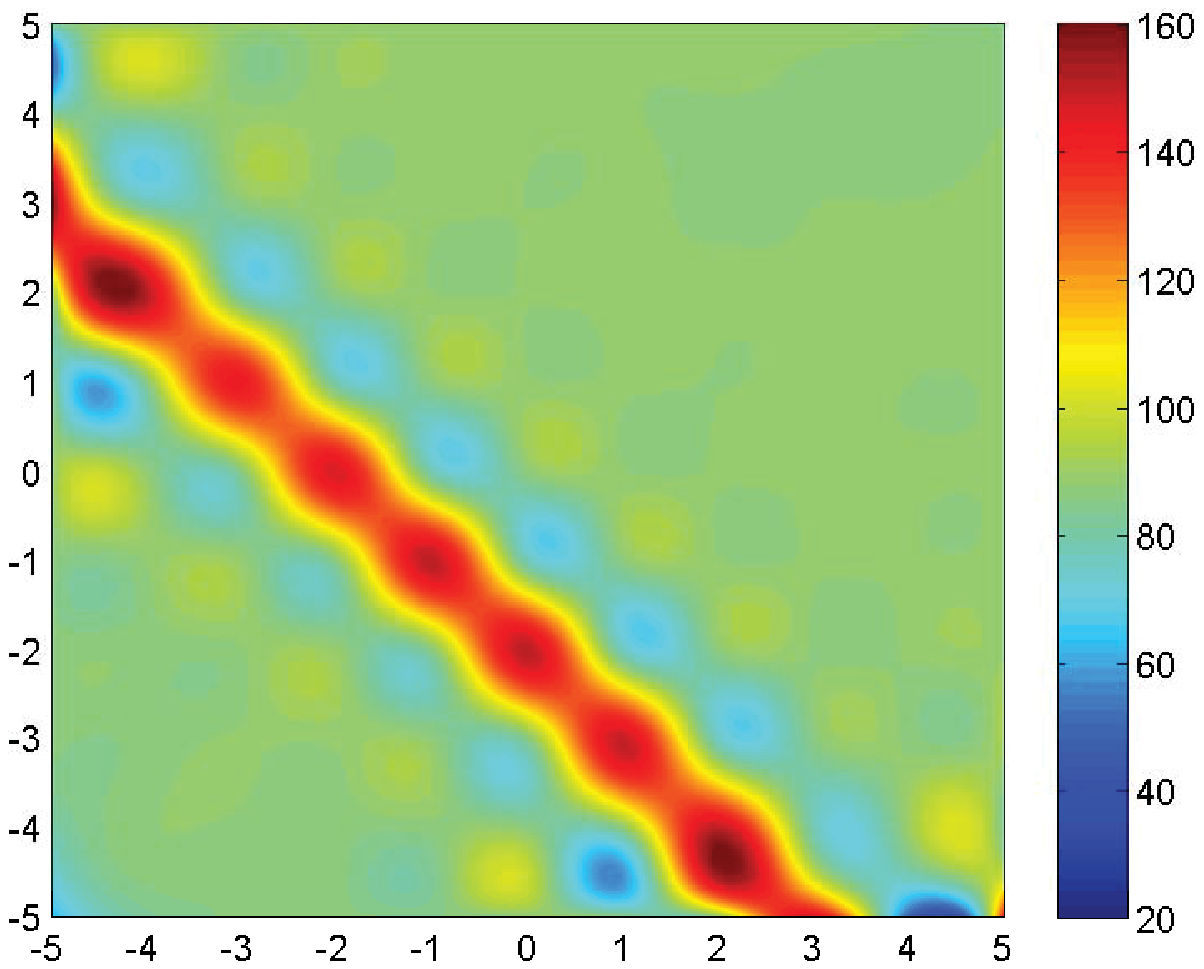}}
\subfigure[]{\label{fig:5:b}\includegraphics[width=2.5in]{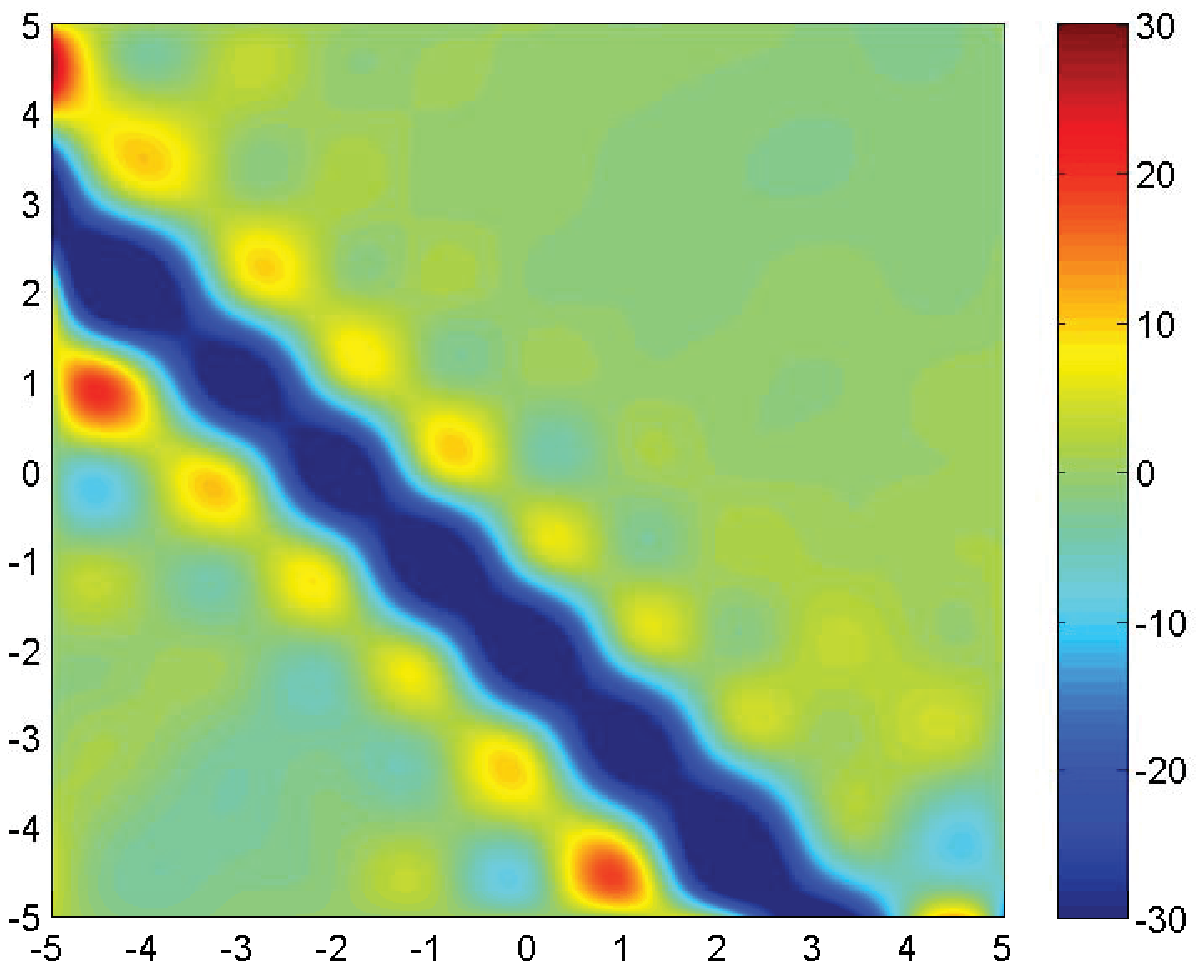}}
\caption{The Cauchy stress component ${\sigma _{yy}}$ (a), ${\sigma _{xz}}$ (b) under tension (color on line).}
\label{fig:5}
\end{figure}

The tensile load is applied along the $y$ direction. In each simulation step, the outermost atoms are moved $0.2A$, and the displacements of other atoms are determined by their distances from the outermost atoms. FIG. \ref{fig:3} shows the configuration and the dislocation distribution after $3600$ picoseconds from the equilibrium configuration. Hence the total length increment is $12A$. FIG. \ref{fig:4} and FIG. \ref{fig:5} shows the strain and stress fields of the cross section.

FIG. \ref{fig:5} shows that the nanowire displays a non-linear response when dislocations appear in the simulation. The dislocation regions are the local maximum points in the strain field, and also by FIG. \ref{fig:5} we can see that the stress field derived by QACC model can exactly represent the mechanical response in the simulation process. The inhomogeneous behavior
of the stress fields are reflected, and the non-local crack propagations are predicted. In addition, because of neglecting the deformation information, traditional Virial stress fields computed by other MD programs are not enough in the micro-nano material characterization.

\subsection{BENDING}

\begin{figure}[ht]
\subfigure[]{\label{fig:6:a}\includegraphics[width=2.5in]{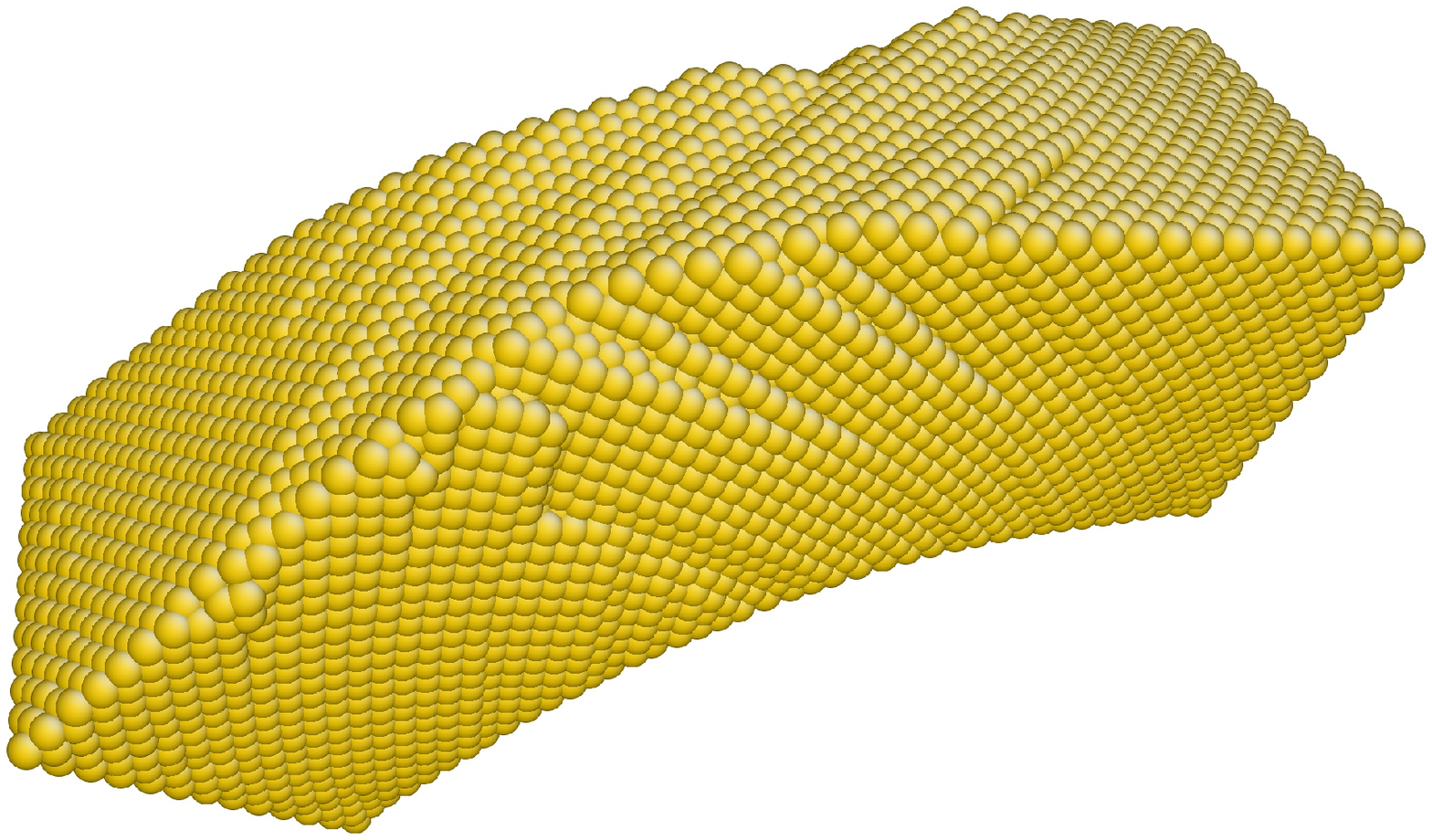}}
\subfigure[]{\label{fig:6:b}\includegraphics[width=2.5in]{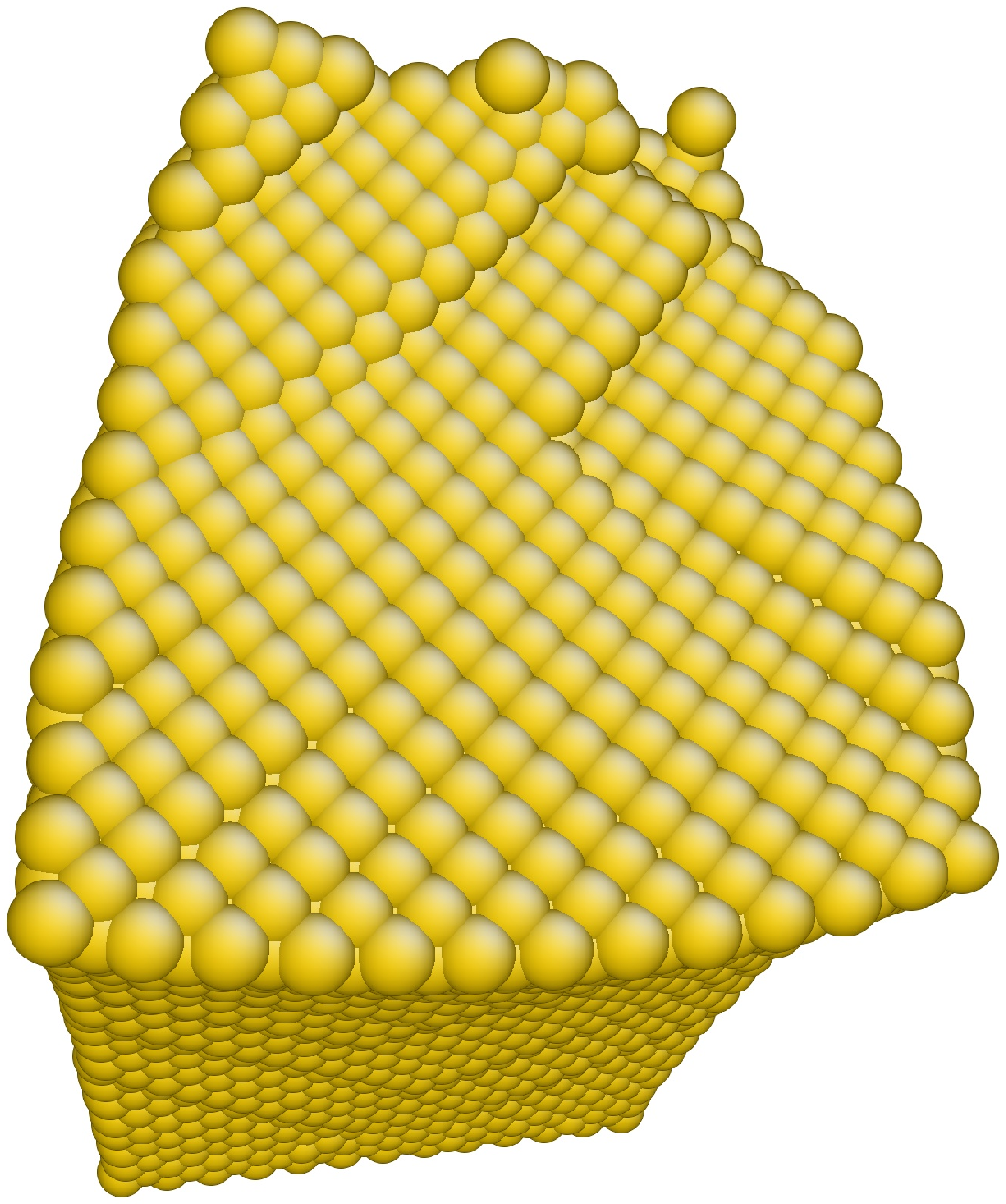}}
\caption{The configuration and the dislocation distribution of the nanowire (a) and the cross section (b) under bending at $1800$ picoseconds (color on line).}
\label{fig:6}
\end{figure}

\begin{figure}[ht]
\subfigure[]{\label{fig:7:a}\includegraphics[width=2.5in]{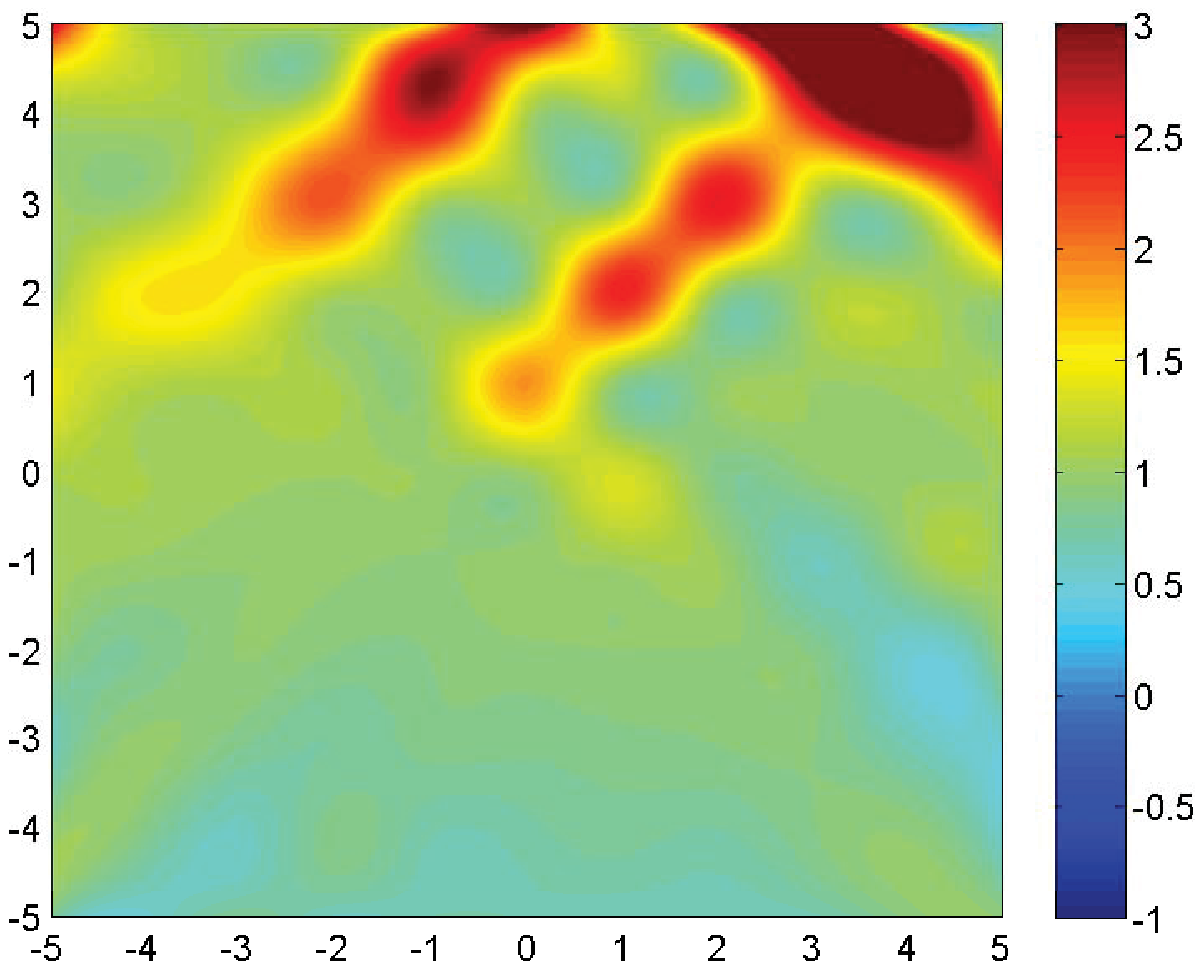}}
\subfigure[]{\label{fig:7:b}\includegraphics[width=2.5in]{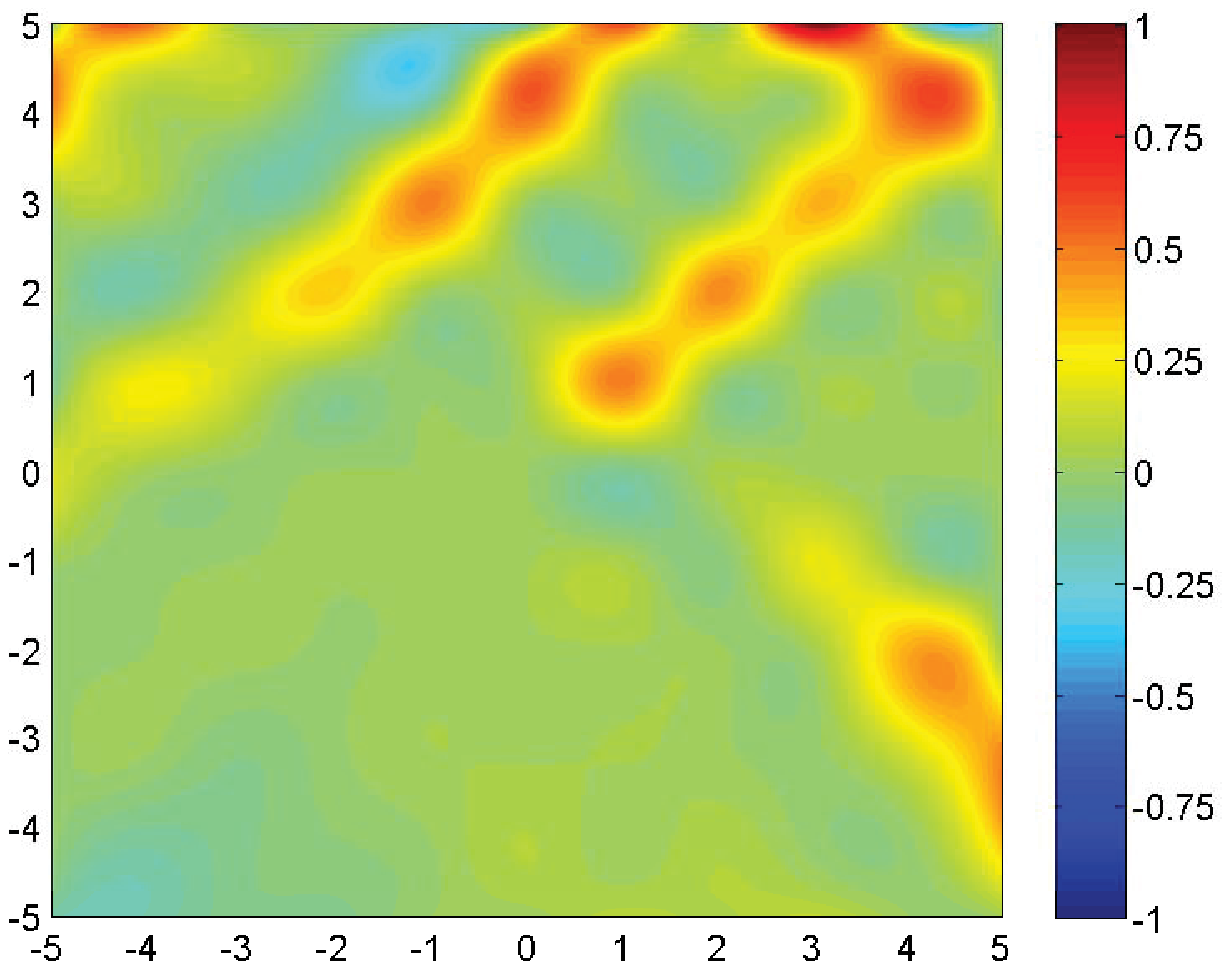}}
\caption{The Cauchy strain component ${E_{yy}}$ (a), ${E_{xz}}$ (b) under bending (color on line).}
\label{fig:7}
\end{figure}

\begin{figure}[ht]
\subfigure[]{\label{fig:8:a}\includegraphics[width=2.5in]{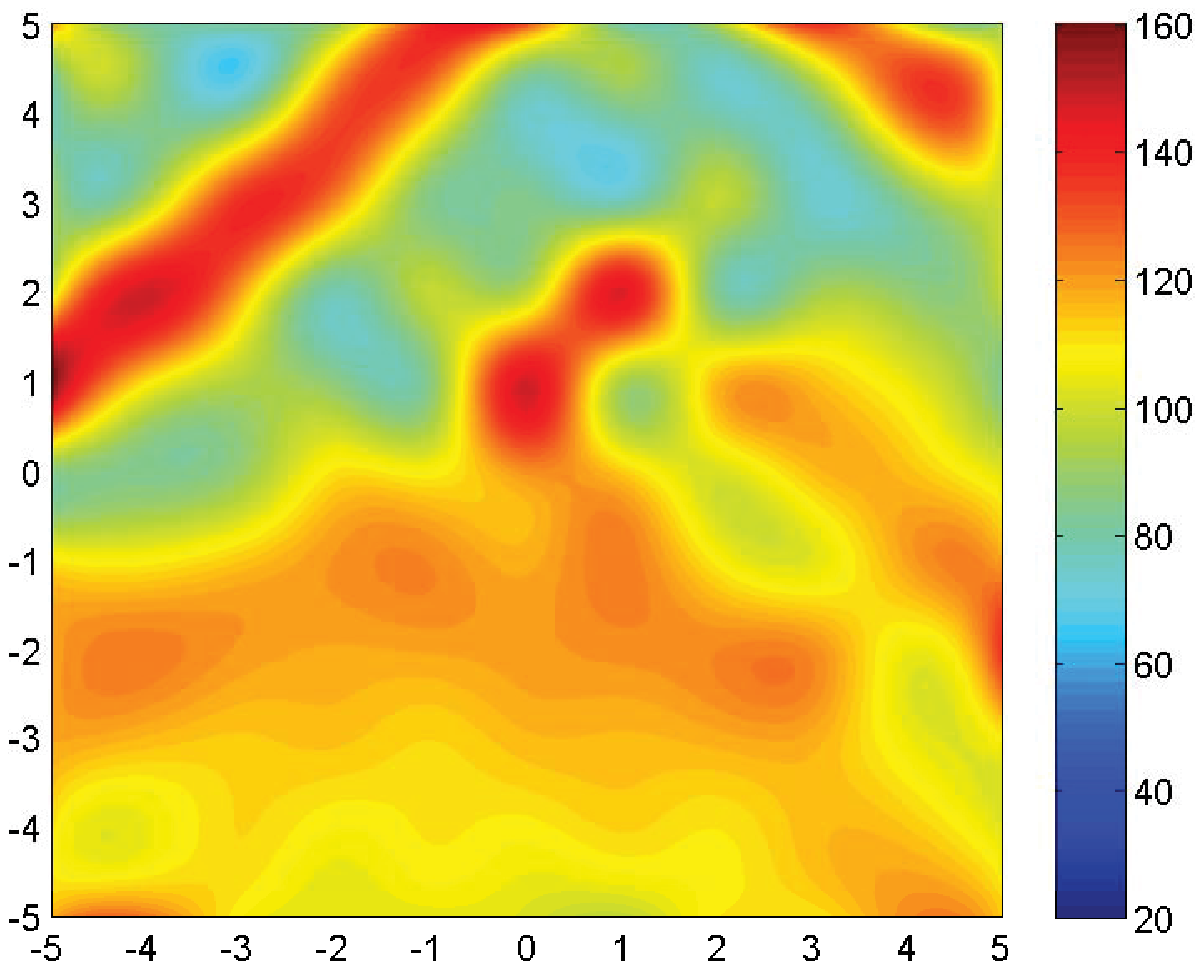}}
\subfigure[]{\label{fig:8:b}\includegraphics[width=2.5in]{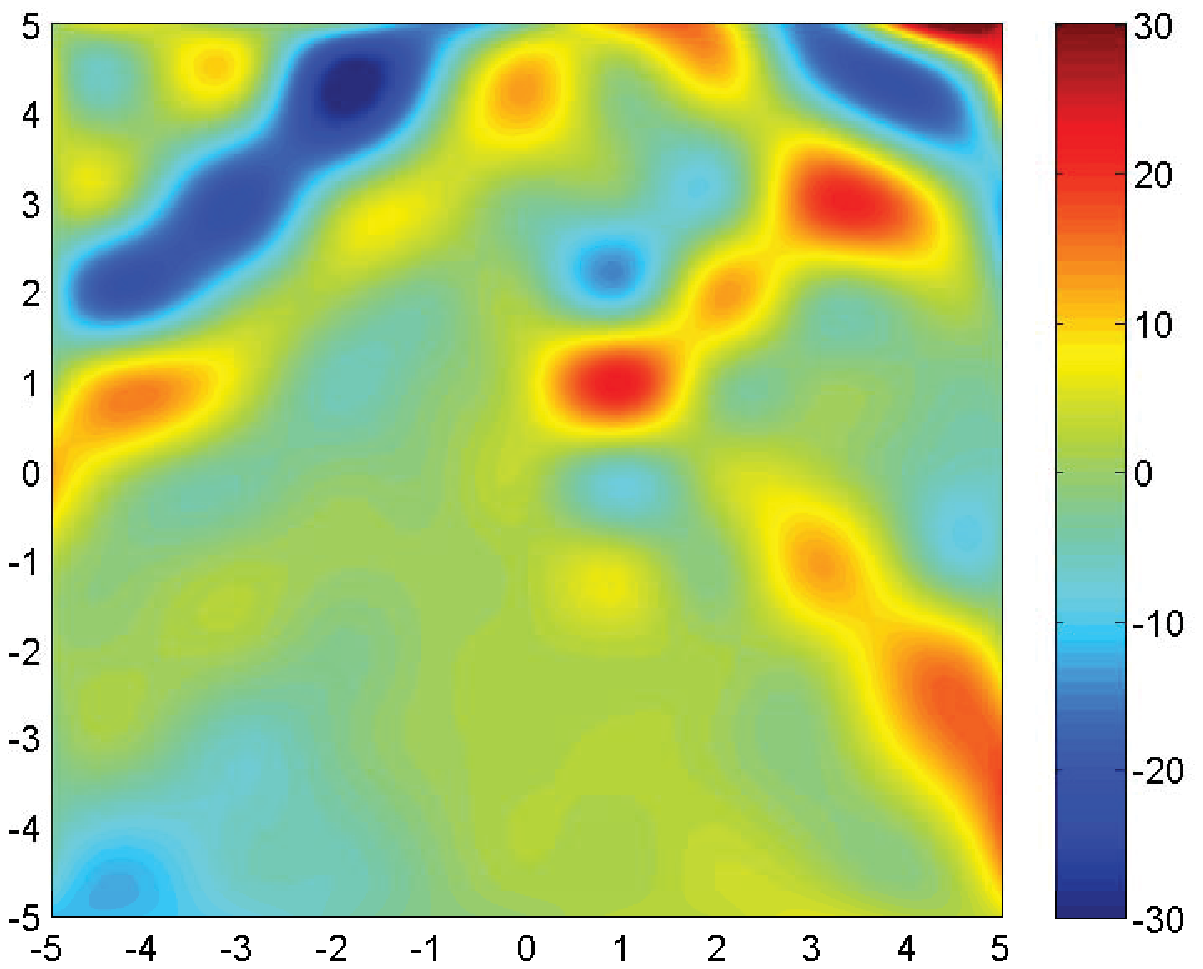}}
\caption{The Cauchy stress component ${\sigma _{yy}}$ (a), ${\sigma _{xz}}$ (b) under bending (color on line).}
\label{fig:8}
\end{figure}

The bending load is applied along the $-z$ direction. In each simulation step, the center of the nanowire is fixed and an angle increment of $1.8$ degrees is exerted. FIG. \ref{fig:6} shows the configuration and the dislocation distribution at a bending angle of $54$ degrees after $1800$ picoseconds from the equilibrium configuration. FIG. \ref{fig:7} and FIG. \ref{fig:8} shows the strain and stress fields of the cross section.

Similarly, the variation of the strain and stress fields is shown clearly, which means that the results given by QACC model are agree with the real physical processes. Note that for different strain rates, and our model gives rational predictions of the non-local crack propagations. As the transferability guaranteed, it may become an available tool in the micro-damage analysis.

\section{V. Conclusions}

A Quantum-Atomic-Continuum-Coupled model has been formulated for the micro-nano material characterization in our paper. And the detailed calculation framework has been built. From the Kohn-Sham equation, the quantum energy density of atomic system has been determined. Deformation gradient, strain and stress tensors and other elastic mechanical quantities have been derived explicitly based on the complex Bravais lattice. To confirm the validity of our model, 3-dimensional numerical simulations have been carried out for single crystal copper nanowires, and reasonable analyses of the strain and stress fields under external loads have been given.

\begin{acknowledgments}

\section{Acknowledgements}

This work is supported by the National Basic Research Program of China (973 Program No. 2012CB025904) , and also supported by the State Key Laboratory of Science and Engineering Computing. The authors gratefully acknowledge help discussions with PhD. Bowen Li and PhD. Yuran Zhang.

\end{acknowledgments}

\appendix

\setcounter{equation}{0}
\renewcommand{\theequation}{A.\arabic{equation}}

\section{APPENDIX}

The first Piola-Kirchhoff stress tensor is denoted by
\begin{eqnarray}
{P_{hK}} = \frac{{\partial {a_r}}}{{\partial {F_{hK}}}} = \frac{1}{V}\sum\limits_{\alpha \in {\Omega _r}} {\frac{{\partial \left\{ {{\eta _\alpha }{E^\alpha } + {k_B}T\sum\limits_{i \in \alpha } {\left[ {{f_i}\ln {f_i} + \left( {1 - {f_i}} \right)\ln \left( {1 - {f_i}} \right)} \right]} } \right\}}}{{\partial {F_{hK}}}}} .
\label{eq:A.1}
\end{eqnarray}
Note that ${E^\alpha } = {\varepsilon ^\alpha } + E_{rep}^\alpha + E_{nucl}^\alpha $, so there are actually four terms in Eq. \eqref{eq:A.1}. For the term $\frac{{\partial {\varepsilon ^\alpha }}}{{\partial {F_{hK}}}}$, we consider $\delta {\varepsilon ^\alpha } = \frac{{\partial {\varepsilon ^\alpha }}}{{\partial {F_{hK}}}}\delta {F_{hK}}\left( X \right)$, and obtain
\begin{eqnarray}
\int_{{\Omega _0}} {\delta {\varepsilon ^\alpha }dV} = \int_{{\Omega _0}} {\frac{{\partial {\varepsilon ^\alpha }}}{{\partial {F_{hK}}}}\delta {F_{hK}}\left( X \right)dV} ,
\label{eq:A.2}
\end{eqnarray}
by integrating both sides on ${\Omega _0}$. On the other hand, we consider
\begin{eqnarray}
\int_{{\Omega _0}} {\delta {\varepsilon ^\alpha }dV} && = \int_{{\Omega _0}} {{{\left( {\frac{{\partial {\varepsilon ^\alpha }}}{{\partial r}}} \right)}^h}\delta {r^h}dV} \nonumber \\
&& = \int_{{\Omega _0}} {{{\left( {\frac{{\partial {\varepsilon ^\alpha }\left( {\int_0^1 {F\left( {{X_\alpha } + s'{R_{\alpha \beta }}} \right){R_{\alpha \beta }}ds'} } \right)}}{{\partial r}}} \right)}^h}\delta \left( {\int_0^1 {{F_{hK}}\left( {{X_\alpha } + s{R_{\alpha \beta }}} \right)R_{\alpha \beta }^Kds} } \right)dV} \nonumber \\
&& = \int_{{\Omega _0}} {\left[ {\int_0^1 {{{\left( {\frac{{\partial {\varepsilon ^\alpha }\left( {\int_0^1 {F\left( {{X_\alpha } + s'{R_{\alpha \beta }}} \right){R_{\alpha \beta }}ds'} } \right)}}{{\partial r}}} \right)}^h}\delta {F_{hK}}\left( {{X_\alpha } + s{R_{\alpha \beta }}} \right)R_{\alpha \beta }^Kds} } \right]dV} ,
\label{eq:A.3}
\end{eqnarray}
where $\delta $ is the Dirac function. Equal Eq. \eqref{eq:A.2} and Eq. \eqref{eq:A.3}, we have
\begin{eqnarray}
\frac{{\partial {\varepsilon ^\alpha }}}{{\partial {F_{hK}}}}\delta {F_{hK}}\left( X \right) = \int_0^1 {{{\left( {\frac{{\partial {\varepsilon ^\alpha }\left( {\int_0^1 {F\left( {{X_\alpha } + s'{R_{\alpha \beta }}} \right){R_{\alpha \beta }}ds'} } \right)}}{{\partial r}}} \right)}^h}\delta {F_{hK}}\left( {{X_\alpha } + s{R_{\alpha \beta }}} \right)R_{\alpha \beta }^Kds} .
\label{eq:A.4}
\end{eqnarray}
for any $\delta F\left( X \right)$. Then for any fixed ${X_\alpha } \in {\Omega _0}$, let
\begin{eqnarray}
\delta {F_{hK}}\left( X \right) = \left\{
\begin{aligned}
0 && other \indent cases , \\
1 && {\left( {h,K} \right) = \left( {m,N} \right)\& X = {X_\alpha }} . \\
\end{aligned}
\right.
\label{eq:A.5}
\end{eqnarray}
we can derive by substituting Eq. \eqref{eq:A.5} into Eq. \eqref{eq:A.4}
\begin{eqnarray}
\frac{{\partial {\varepsilon ^\alpha }}}{{\partial {F_{hK}}}} = \int_0^1 {\left( {\frac{{\partial {\varepsilon ^\alpha }\left( r \right)}}{{\partial r}}} \right)_{r = \int_0^1 {F\left( {X + \left( {s' - s} \right){R_{\alpha \beta }}} \right){R_{\alpha \beta }}ds'} }^hR_{\alpha \beta }^Kds} .
\label{eq:A.6}
\end{eqnarray}
With the help of Hellmann-Feynman theorem, we can show
\begin{eqnarray}
\frac{{\partial {\varepsilon _\alpha }}}{{\partial r}} = && \sum\limits_{i \in \alpha } { \langle \frac{{\partial {\phi _i}}}{{\partial r}}| - \frac{1}{2}\nabla _i^2 + {V_{eff}}\left( {{\rho _\alpha }} \right)|{\phi _i} \rangle } + \sum\limits_{i \in \alpha } { \langle {\phi _i}| - \frac{1}{2}\nabla _i^2 + {V_{eff}}\left( {{\rho _\alpha }} \right)|\frac{{\partial {\phi _i}}}{{\partial r}} \rangle } \nonumber \\
&& + \sum\limits_{i \in \alpha } {\langle {{\phi _i}} | \frac{{\partial \left( { - \frac{1}{2}\nabla _i^2 + {V_{eff}}\left( {{\rho _\alpha }} \right)} \right)}}{{\partial r}} | {{\phi _i}} \rangle } \nonumber \\
= && \sum\limits_{i \in \alpha } { \langle \frac{{\partial {\phi _i}}}{{\partial r}} | {\varepsilon _i}{\phi _i} \rangle } + \sum\limits_{i \in \alpha } { \langle {\varepsilon _i}{\phi _i} | \frac{{\partial {\phi _i}}}{{\partial r}} \rangle } + \sum\limits_{i \in \alpha } { \langle {\phi _i} | \frac{{\partial \left( { - \frac{1}{2}\nabla _i^2 + {V_{eff}}\left( {{\rho _\alpha }} \right)} \right)}}{{\partial r}} | {\phi _i} \rangle } \nonumber \\
= && \sum\limits_{i \in \alpha } {{\varepsilon _i}\frac{{\partial \langle {\phi _i} | {\phi _i} \rangle }}{{\partial r}}} + \sum\limits_{i \in \alpha } { \langle {\phi _i} | \frac{{\partial \left( { - \frac{1}{2}\nabla _i^2 + {V_{eff}}\left( {{\rho _\alpha }} \right)} \right)}}{{\partial r}} | {\phi _i} \rangle } \nonumber \\
= && \sum\limits_{i \in \alpha } { \langle {\phi _i} | \frac{{\partial \left( { - \frac{1}{2}\nabla _i^2 + {V_{eff}}\left( {{\rho _\alpha }} \right)} \right)}}{{\partial r}} | {\phi _i} \rangle } .
\label{eq:A.7}
\end{eqnarray}
Now the first term of Eq. \eqref{eq:A.1} is completely determined. And by the same skill, we can determine the first three terms. For the last term of Eq. \eqref{eq:A.1}, we write
\begin{eqnarray}
\sum\limits_i {{f_i}} = \sum\limits_i {\frac{1}{{\exp \left( {{{{\varepsilon _i} - {\varepsilon _f}} \mathord{\left/ {\vphantom {{{\varepsilon _i} - {\varepsilon _f}} {{k_B}T}}} \right. \kern-\nulldelimiterspace} {{k_B}T}}} \right) + 1}}} = {n_{elec}} ,
\label{eq:A.8}
\end{eqnarray}
by differentiating both sides of Eq. \eqref{eq:A.8} with respect to ${\varepsilon _i}$, we obtain $\frac{{\partial {\varepsilon _f}}}{{\partial {\varepsilon _i}}} = \frac{{{g_i}}}{{\sum\limits_m {{g_m}} }}$ and ${g_i} = - \frac{{\exp \left( {{{{\varepsilon _i} - {\varepsilon _f}} \mathord{\left/ {\vphantom {{{\varepsilon _i} - {\varepsilon _f}} {{k_B}T}}} \right. \kern-\nulldelimiterspace} {{k_B}T}}} \right)}}{{{{\left( {\exp \left( {{{{\varepsilon _i} - {\varepsilon _f}} \mathord{\left/ {\vphantom {{{\varepsilon _i} - {\varepsilon _f}} {{k_B}T}}} \right. \kern-\nulldelimiterspace} {{k_B}T}}} \right) + 1} \right)}^2}}}$. Thus
\begin{eqnarray}
\frac{{\partial {f_i}}}{{\partial {\varepsilon _n}}} = \frac{{\partial {f_i}}}{{\partial {\varepsilon _i}}}\frac{{\partial {\varepsilon _i}}}{{\partial {\varepsilon _n}}} + \frac{{\partial {f_i}}}{{\partial {\varepsilon _f}}}\frac{{\partial {\varepsilon _f}}}{{\partial {\varepsilon _n}}} = \frac{{{g_i}}}{{{k_B}T}}\left( {{\delta _{i,n}} - \frac{{{g_n}}}{{\sum\limits_m {{g_m}} }}} \right) ,
\label{eq:A.9}
\end{eqnarray}
and
\begin{eqnarray}
\frac{1}{{{V_r}}} && \sum\limits_{\alpha \in {\Omega _r}} {{k_B}T\frac{{\sum\limits_{i \in \alpha } {\partial \left[ {{f_i}\ln {f_i} + \left( {1 - {f_i}} \right)\ln \left( {1 - {f_i}} \right)} \right]} }}{{\partial {F_{hK}}}}} \nonumber \\
&& = \frac{1}{{{V_r}}}\sum\limits_{i \in {\Omega _r}} {\sum\limits_n {{k_B}T\frac{{\partial \left[ {{f_i}\ln {f_i} + \left( {1 - {f_i}} \right)\ln \left( {1 - {f_i}} \right)} \right]}}{{\partial {\varepsilon _n}}}\frac{{\partial {\varepsilon _n}}}{{\partial {F_{hK}}}}} } \nonumber \\
&& = \frac{1}{{{V_r}}}\sum\limits_{i \in {\Omega _r}} {\sum\limits_n {{k_B}T\ln \frac{{{f_i}}}{{1 - {f_i}}}\frac{{\partial {f_i}}}{{\partial {\varepsilon _n}}}\frac{{\partial {\varepsilon _n}}}{{\partial {F_{hK}}}}} } \nonumber \\
&& = - \frac{1}{{{V_r}}}\sum\limits_{i \in {\Omega _r}} {\left[ {\frac{{{g_i}}}{{{k_B}T}}\left( {{\varepsilon _i} - \frac{{\sum\limits_m {{g_m}{\varepsilon _m}} }}{{\sum\limits_m {{g_m}} }}} \right)} \right]\frac{{\partial {\varepsilon _i}}}{{\partial {F_{hK}}}}} .
\label{eq:A.10}
\end{eqnarray}
As $\frac{{\partial {\varepsilon _i}}}{{\partial {F_{hK}}}}$ is shown previously, we finally derive
\begin{eqnarray}
{P_{hK}} = && \frac{1}{V}\sum\limits_{\alpha \in {\Omega _r}} {\left\{ {\sum\limits_{\beta \ne \alpha } {{\eta _\alpha }\left[ {\int_0^1 {\left( {\frac{{\partial \left( {{\varepsilon ^\alpha }\left( r \right) + E_{rep}^\alpha \left( \rho \right) + {Z_\alpha }{Z_\beta }{r^{ - 1}}} \right)}}{{2\partial r}}} \right)_{r = \int_0^1 {F\left( {X + \left( {s' - s} \right){R_{\alpha \beta }}} \right){R_{\alpha \beta }}ds'} }^hR_{\alpha \beta }^Kds} } \right]} } \right.} \nonumber \\
&& \left. { - \sum\limits_{i \in \alpha } {\left[ {\frac{{{g_i}}}{{{k_B}T}}\left( {{\varepsilon _i} - \frac{{\sum\limits_m {{g_m}{\varepsilon _m}} }}{{\sum\limits_m {{g_m}} }}} \right)\int_0^1 {\left( {\frac{{\partial {\varepsilon _i}\left( r \right)}}{{\partial r}}} \right)_{r = \int_0^1 {F\left( {X + \left( {s' - s} \right){R_{\alpha \beta }}} \right){R_{\alpha \beta }}ds'} }^hR_{\alpha \beta }^Kds} } \right]} } \right\} .
\label{eq:A.11}
\end{eqnarray}
Similarly
\begin{eqnarray}
{D_{iJkL}} = && \frac{{{\partial ^2}{a_r}}}{{\partial {F_{iJ}}\partial {F_{kL}}}} = \frac{{\partial {P_{iJ}}}}{{\partial {F_{kL}}}} \nonumber \\
= && \frac{1}{V}\sum\limits_{\alpha \in {\Omega _r}} {\left\{ {\sum\limits_{\beta \ne \alpha } {{\eta _\alpha }\left[ {\int_0^1 {\int_0^1 {\left( {\frac{{{\partial ^2}\left( {{\varepsilon ^\alpha }\left( r \right) + E_{rep}^\alpha \left( \rho \right) + {Z_\alpha }{Z_\beta }{r^{ - 1}}} \right)}}{{2\partial {r^2}}}} \right)_{r = \int_0^1 {F\left( {X + \left( {s'' - s' - s} \right){R_{\alpha \beta }}} \right){R_{\alpha \beta }}ds''} }^{i,k}R_{\alpha \beta }^JR_{\alpha \beta }^Lds'} ds} } \right]} } \right.} \nonumber \\
&& \left. { - \sum\limits_{i \in \alpha } {\left[ {\frac{{{g_i}}}{{{k_B}T}}\left( {{\varepsilon _i} - \frac{{\sum\limits_m {{g_m}{\varepsilon _m}} }}{{\sum\limits_m {{g_m}} }}} \right)\int_0^1 {\int_0^1 {\left( {\frac{{{\partial ^2}{\varepsilon _i}\left( r \right)}}{{\partial {r^2}}}} \right)_{r = \int_0^1 {F\left( {X + \left( {s'' - s' - s} \right){R_{\alpha \beta }}} \right){R_{\alpha \beta }}ds''} }^{i,k}R_{\alpha \beta }^JR_{\alpha \beta }^Lds'} ds} } \right]} } \right\} .
\label{eq:A.12}
\end{eqnarray}
Note that the second Piola-Kirchhoff stress tensor is ${S_{IJ}} = 2\frac{{\partial {a_r}}}{{\partial {E_{IJ}}}} = \sum\limits_h {F_{Ih}^{ - 1}{P_{hJ}}}$, and the elastic tensor is determined by the chain rule
\begin{eqnarray}
{C_{IJKL}} && = 4\frac{{{\partial ^2}{a_r}}}{{\partial {E_{IJ}}\partial {E_{KL}}}} = 2\frac{{\partial {S_{IJ}}}}{{\partial {E_{KL}}}} = 2\sum\limits_h {\frac{{\partial {{\left( {{F^{ - 1}}} \right)}_{Ih}}{P_{hJ}}}}{{\partial {E_{KL}}}}} \nonumber \\
&& = 2\sum\limits_{h,m,N} {\left[ {\frac{{\partial {{\left( {{F^{ - 1}}} \right)}_{Ih}}}}{{\partial {F_{mN}}}}{P_{hJ}} + {{\left( {{F^{ - 1}}} \right)}_{Ih}}\frac{{\partial {P_{hJ}}}}{{\partial {F_{mN}}}}} \right]\frac{{\partial {F_{mN}}}}{{\partial {E_{KL}}}}} \nonumber \\
&& = 2\sum\limits_{h,m,N} {\left[ {\frac{{\partial {{\left( {{F^{ - 1}}} \right)}_{Ih}}}}{{\partial {F_{mN}}}}{P_{hJ}} + {{\left( {{F^{ - 1}}} \right)}_{Ih}}{D_{hJmN}}} \right]\frac{{\partial {F_{mN}}}}{{\partial {E_{KL}}}}} ,
\label{eq:A.13}
\end{eqnarray}
where $\frac{{\partial F}}{{\partial E}} = {\left( {\frac{{\partial E}}{{\partial F}}} \right)^{ - 1}}$. Thus we have given all the detailed derivations in Section III.


\begin{thebibliography}{99}

\bibitem{1} C. C. Fu, J. D. Torre, F. Willaime, J. L. Bocquet, and A. Barbu, Nat. Mater. {\bf4}, 68 (2005).
\bibitem{2} R. E. Rudd and J. Q. Broughton, Phys. Rev. B {\bf58}, R5893 (1998).
\bibitem{3} M. Elstner, D. Porezag, G. Jungnickel, J. Elsner, M. Haugk, Th. Frauenheim, S. Suhai, and G. Seifert, Phys. Rev. B {\bf58}, 7260 (1998).
\bibitem{4} D. W. Brenner, O. A. Shenderova, J. A. Harrison, S. J. Stuart, B. Ni, and S. B. Sinnott, J. Phys. Condens. Matter {\bf14}, 783 (2002).
\bibitem{5} A. J. Cohen, P. Mori-S\'{a}nchez, and W. Yang, Chem. Rev. {\bf112}, 289 (2011).
\bibitem{6} M. van Faassen, P. L. de Boeij, R. van Leeuwen, J. A. Berger, and J. G. Snijders, Phys. Rev. Lett. {\bf88}, 186401 (2002).
\bibitem{7} E. B. Tadmor, M. Ortiz, and R. Phillips, Philos. Mag. A {\bf73}, 1529 (1996).
\bibitem{8} G. J. Wagner, and W. K. Liu, J. Comput. Phys. {\bf190}, 249 (2003).
\bibitem{9} E. Weinan, B. Engquist, and Z. Huang, Phys. Rev. B {\bf67}, 092101 (2003).
\bibitem{10} M. Dobson, and M. Luskin, ESAIM. Math. Model. Numer. Anal. {\bf43}, 591 (2009).
\bibitem{11} S. A. Brenner, and L. R. cott, \emph{The Mathematical Theory of Finite Element Methods} (Springer, NY, 2008).
\bibitem{12} Z. P. Bazant, and M. Jir\'{a}sek, J. Eng. Mech. {\bf128}, 1119 (2002).
\bibitem{13} D. Marx, and J. Hutter, \emph{Ab Initio Molecular Dynamics} (Cambridge University Press, NY, 2009).
\bibitem{14} M. F. Kanninen, and C. H. Popelar, \emph{Advanced Fracture Mechanics} (Oxford University Press, NY, 1985).
\bibitem{15} J. P. Perdew, K. Burke, and M. Ernzerhof, Phys. Rev. Lett. {\bf77}, 3865 (1996).
\bibitem{16} E. B. Tadmor, G. S. Smith, N. Bernstein, and E. Kaxiras, Phys. Rev. B {\bf59}, 235 (1999).
\bibitem{17} R. S. Barsoum, Int. J. Numer. Method. Eng. {\bf10}, 25 (1976).
\bibitem{18} B. Li, J. Cui, X. Tian, X. Yu, and M. Xiang, Comput. Mater. Sci. http://dx.doi.org/10.1016/j.commatsci.2014.02.002, (2014).
\bibitem{19} M. Levy, and J. P. Perdew, Phys. Rev. A {\bf32}, 2010 (1985).
\bibitem{20} M. Zhou, Proc. Roy. Soc. London Ser. A {\bf459}, 2347 (2003).
\bibitem{21} M. J. Frisch, et al., Gaussian 09, Revision A. 02 (Gaussian, Inc., Wallingford CT, 2009).
\bibitem{22} P. J. Hay, W. R. Wadt, J. Chem. Phys. {\bf82}, 299 (1985).
\bibitem{23} J. Tao, J. P. Perdew, V. N. Staroverov, and G. E. Scuseria, Phys. Rev. Lett. {\bf91}, 146401 (2003).

\end{thebibliography}
\end{document}